ARTICLE

# Hot Carrier Organic Solar Cells

Priya Viji,[a] Constantin Tormann,[a] Clemens Göhler [a] and Martijn Kemerink *[a]

Hot-carrier solar cells use the photon excess energy, that is, the energy exceeding the absorber bandgap, to do additional work. These devices have the potential to beat the upper limit for the photovoltaic power conversion efficiency set by near-equilibrium thermodynamics. However, since their conceptual inception in 1982, no experimental realization that works under normal operational conditions has been demonstrated, mostly due to the fast thermalization of photo-generated charges in typical semiconductor materials. Here, we use noise spectroscopy in combination with numerical modelling to show that common bulk heterojunction organic solar cells actually work as hot-carrier devices. Due to static energetic disorder, thermalization of photo-generated electrons and holes in the global density of states is slow compared to the charge carrier lifetime, leading to thermal populations of localized charge carriers that have an electronic temperature exceeding the lattice temperature. Since charge extraction takes place in a high-lying, narrow energy window around the transport energy, the latter takes the role of an energy filter. For common disorder values, this leads to enhancements in open circuit voltage of up to ~0.2 V. We show that this enhancement can be understood as a thermovoltage that is proportional to the temperature difference between the lattice and the charge populations and that comes on top of the near-equilibrium quasi-Fermi level splitting.

## Broader context

One of the major efficiency limiting factors in conventional solar cells are thermalization losses, which refers to all photon energies in excess of the absorber bandgap being lost due to photogenerated charges quickly reaching thermal equilibrium. A conceptually simple way to harvest this excess energy was proposed by Ross and Nozik in 1982, and relies on (i) slowing down thermalization to achieve 'hot' populations of electrons and holes and (ii) selectively extracting charges from these populations at energies exceeding the bandgap. Making this concept work under practical conditions has proven an unsurmountable hurdle, mostly due to the fast thermalization of photo-generated charges in typical semiconductor materials like silicon. Here, we use noise spectroscopy in combination with numerical modelling to show that the electronic temperature of photogenerated charges in organic solar cells under standard operational conditions far exceeds the lattice temperature and that these devices, in stark contrast to all other (inorganic and hybrid) solar cells, actually work as hot-carrier devices as envisioned by Ross and Nozik. Our results show a new way to realize hot-carrier solar cells and indicate a need to rethink how energy losses in organic solar cells are understood and, accordingly, can be minimized.

## Introduction

In their seminal 1982 paper, Ross and Nozik introduced the concept of harnessing excess energy from photo-absorption that could mitigate thermalization losses and, therefore, surpass the detailed balanced limit.[1] They coined the term "hot-carrier" solar cell (HCSC) for a device in which photo-excited charges thermally equilibrate among themselves but are extracted before reaching an equilibrium with the lattice. These HCSCs can potentially achieve power conversion efficiencies up to 66% for a single bandgap absorber, which is to be contrasted with the near-equilibrium Shockley-Queisser limit of about 30% for a single junction and 68% for a multi-junction solar cell consisting of an infinite number of layers. Despite the conceptualization almost half a century ago, experimental realization of such a device that works around room temperature and with reasonable illumination intensities is essentially non-existent.[2] The main difficulties are the fast, ~1 ps, time scale of thermalization of photo-generated charges by phonon emission in typical inorganic semiconductors like silicon, in combination with the difficulty to construct an efficient energy-selective contact.[3–5] The crucial role of the latter is to selectively extract hot charges while blocking cold charges.

[a] Institute for Molecular Systems Engineering and Advanced Materials, Heidelberg University, Im Neuenheimer Feld 225, 69120 Heidelberg, Germany.
† Supplementary Information (SI) available. See DOI: 10.1039/x0xx00000x

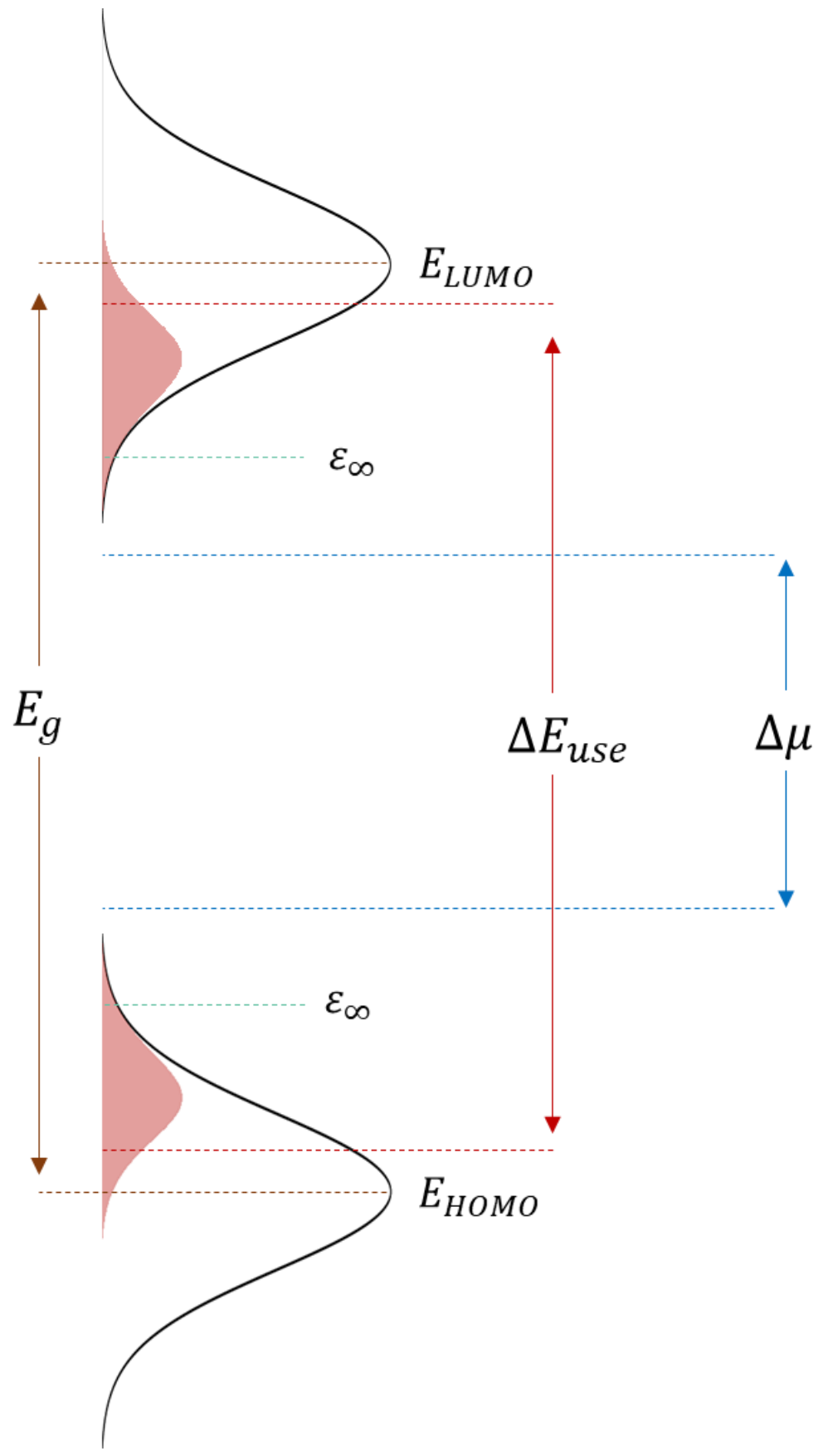

Fig. 1 Schematics of a working hot-carrier organic solar cell. An organic hot-carrier solar cell with disorder-broadened Gaussian DOS (black lines) and non-thermalized density of occupied states (shaded regions, blown up for visibility) with $E_\infty$ the equilibrium energy, $\Delta E_{use}$ the difference in energy between extraction pathway for holes and electrons and $\Delta\mu$ the quasi-Fermi level splitting, which determines the radiative $V_{OC}$ limit.

For a device working as described above and as illustrated in Fig. 1, from which 'hot' charges, i.e., charges that are not thermalized to bandedge, are extracted via a utilization pathway at energy $\Delta E_{use}$, which sits well above the semiconductor bandgap (and absorption onset) $E_{gap}$, Ross and Nozik derived for the open-circuit voltage $V_{OC}$

$$eV_{OC} = \Delta\mu\frac{T_{latt}}{T_{el}} + \Delta E_{use}\left(1 - \frac{T_{latt}}{T_{el}}\right) \qquad (1)$$

When the electronic temperature $T_{el}$ equals the lattice temperature $T_{latt}$, the HCSC converges to a conventional solar cell with $V_{OC}$ governed by the splitting of the electron and hole quasi-Fermi levels, $\Delta\mu$.

The focus of the HCSC community's search for a suitable absorber has predominantly centered on inorganic materials such as GaAs,[2] halide perovskites,[6,7] and hybrid perovskites.[8] These materials have shown hot carrier cooling lifetimes in the order of a few 100 picoseconds to a few nanoseconds, which is slow in the context of inorganic semiconductors but still insufficient in comparison to competing processes. Despite much longer thermalization timescales of up to tens of µs,[9] organic semiconductors were, somewhat surprisingly, never considered for HCSC applications. This could stem from the apparent success of near-equilibrium models to explain a range of experiments, most notably the value of the open circuit voltage and the overall shape of the current-voltage characteristics.[10–12] In addition, energetic disorder in organics is generally understood to be an additional source of energy losses, running counter to the goal of HCSC: full thermalization in a Gaussian density of states (DOS) with a typical width $\sigma_{DOS} =$ 50-90 meV, would amount to energy losses of 0.1-0.3 eV.[13,14]

Although transient absorption spectroscopy can be employed to determine carrier temperatures and decay times in perovskite-based systems,[7,8,15–17] determining the electronic temperature directly from the shape of transient or static optical spectra is impossible in organics due to broadening by static disorder.[18] Nevertheless, experimental signatures of non-thermalized charges in organic solar cells have been seen through time-dependent mobilities , faster-than-equilibrium extraction,[14] and non-thermalized populations of charge transfer (CT) states under steady-state illumination.[19] Moreover, it was argued that the reciprocity relations, which underlie the near-equilibrium treatment of $V_{OC}$ may not hold in organic solar cells.[20]

To avoid confusion, it must be pointed out that thermalization in energetically disordered organic semiconductors, that is, the typical materials used to make state-of-the-art organic photovoltaic (OPV) devices,[21,22] occurs as a two-step process. The first is a fast, mostly onsite, thermalization to the lattice temperature by coupling to molecular vibrations. This is equivalent to cooling by phonon emission in inorganic semiconductors and produces, in a timeframe of ~0.1 ps, a localized polaron.[23,24] Despite being 'locally cold', this polaron is typically not created at the equilibrium energy of the global density of localized states but much above it. For a Gaussian DOS and low charge carrier densities, as typical for good OPV devices, the equilibrium energy sits at $\varepsilon_{eq} = -\sigma_{DOS}^2/k_BT$ while charges are, on average, excited at the DOS center at $\varepsilon = 0$. It is the 'global' thermalization process, by hopping through intermediate sites, towards this equilibrium energy that is slow and is one of the crucial ingredients that make general OPV to hot carrier devices.[25,26]

In this Letter, we utilize Johnson thermometry through cross-correlated current noise spectroscopy to measure the temperature of the charge carrier populations in two representative bulk heterojunction OPV systems, PM6:Y6 and P3HT:PCBM, under operational conditions. In stark contrast to their inorganic counterpart, exemplified by a commercial silicon PV device, the charge distributions in the organic solar cells are almost twice as hot as the lattice. We confirm our experimental results by performing kinetic Monte Carlo (kMC) simulations of typical OPV devices. The simulations quantitatively confirm that the static disorder in the organic material causes the high electronic temperature and concomitantly, the enhanced noise signals. We then connect this finding to the Ross-Nozik model and demonstrate that the open-circuit voltage can be described by Eq. (1), using independently determined parameters.

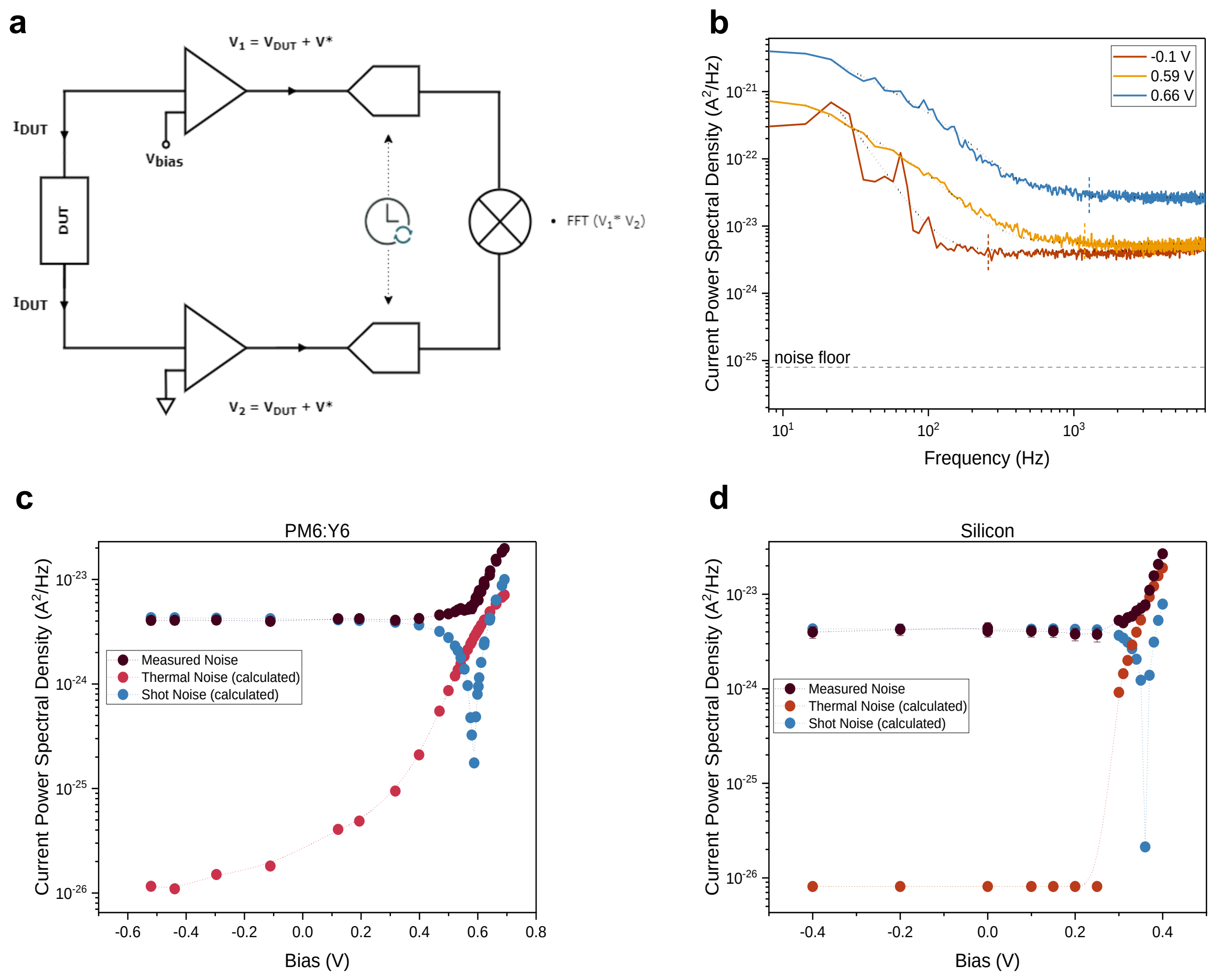

**Fig. 2 Noise spectroscopy setup and measurements.** (**a**) Cross-correlation noise spectroscopy setup. The two terminals of the device under test (DUT) are connected to the inputs of two synchronized lock-in amplifiers through transimpedance amplifiers. The lock-in outputs are cross-correlated to suppress instrument noise. (**b**) Noise spectra from an illuminated PM6:Y6 solar cell at room temperature for different biases. The dashed vertical line indicates the beginning of the white spectra, and the black dotted line is a fit to Eq. 2. (**c**)&(**d**) Noise after cross-talk correction vs bias voltage for sub-1 Sun illuminated PM6:Y6 and silicon PV devices, respectively, with thermal and shot noise values calculated assuming the electronic temperature equals the lattice temperature and a Fano factor $F = 1$.

## Results and discussion

We used cross-correlated current noise spectroscopy to measure the thermal noise and, thereby, the temperatures associated with the electronic charge distributions in the device under test (DUT) using the setup shown in Fig. 2a. The setup and methodology, which corrects for noise arising from cross-talk between the transimpedance amplifiers over the DUT resistance, were tested on commercial resistors and doped organic thin films, as discussed in Supplementary Text S1, S2 and S5. Over a wide resistance range, electronic temperatures equal to the lattice temperature were found.

Experiments were performed on two different, prototypical organic bulk heterojunction systems, the classical polymer-fullerene system P3HT:PCBM and the state-of-the-art PM6:Y6 system. In addition, a commercial inorganic silicon photodiode was tested. Exemplary measurement data of an organic PV device are shown in Fig. 2b and comprise of 1/f-noise, characterized by an exponent $\alpha$, shot noise, that is proportional to the product of the current $I$ and Fano factor $F$, in addition to the thermal (Johnson) noise that depends on the electronic temperature $T_{el}$ and device resistance $R$.

$$S_I(f) = \frac{A}{f^\alpha} + 2qIF + \frac{4k_BT_{el}}{R} \qquad (2)$$

Since the noise of interest, i.e., the thermal noise, has a white spectrum, we only analyze data for which the 1/f-noise is suppressed, e.g. in Fig. 2b that is from about 200 Hz onwards for the -0.1 V measurement.

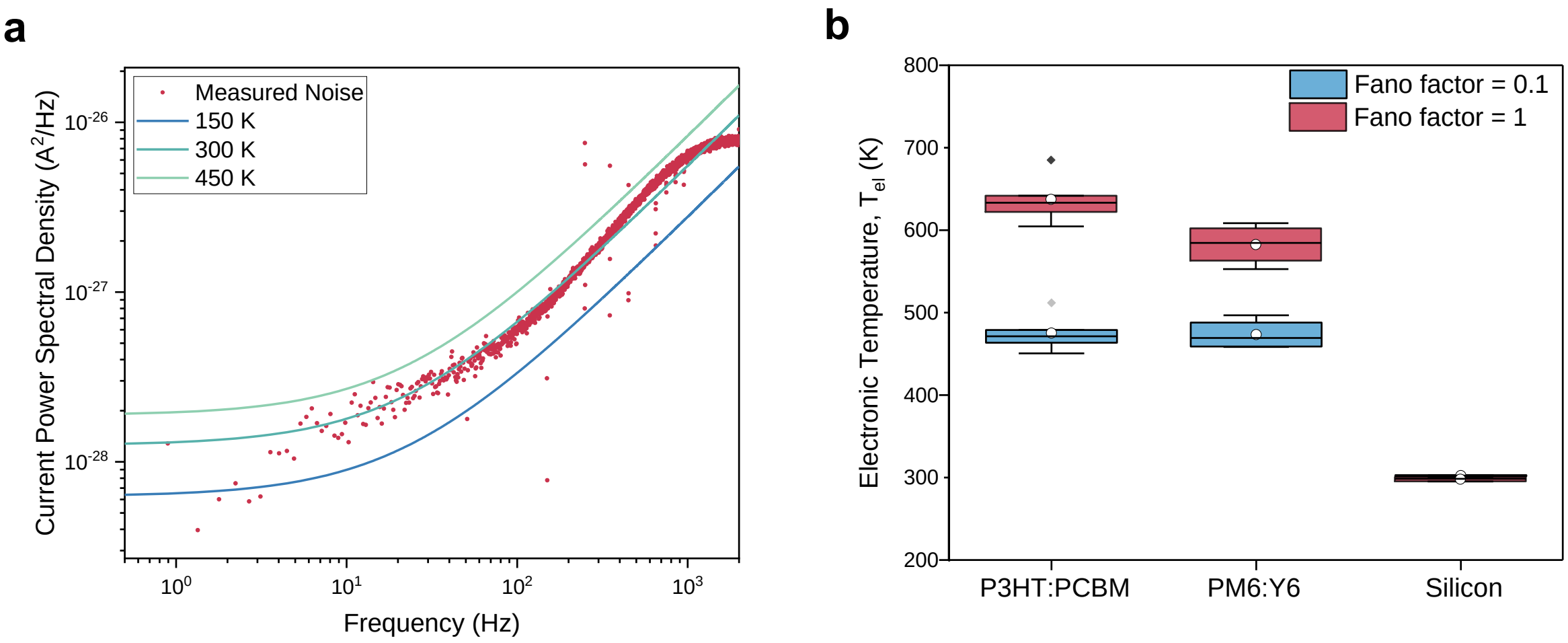

**Fig. 3 Electronic temperature in dark and under illumination.** (**a**) Noise spectra of a PM6:Y6 solar cell in the dark at room temperature and $V = 0$. The colored lines are the predicted thermal noise with different electronic temperatures. A temperature of ~300 K best fits the noise spectra. (**b**) Extracted electronic temperatures at open-circuit conditions for different solar cells under sub-1 Sun illumination intensity at room temperature.

Noise measurements of illuminated solar cells were taken at different biasing voltages, including at short- and open-circuit; Fig. 2c,d plots the resulting plateau values (black symbols), along with estimates of the shot and thermal noise, assuming $F = 1$, $T_{el} = T_{latt}$ and calculating $R$ from the slope of the IV-curve, i.e. $R(V) = (dI(V)/dV)^{-1}$. Focussing on the region around open-circuit, where the shot noise contribution is negligible, highlights a marked difference between the organic (Fig. 2c) and inorganic (Fig. 2d) devices. While for the silicon PV cell, the measured noise coincides with the estimated noise, the PM6:Y6 solar cell shows a significantly higher (Johnson) noise than expected for electronic distributions in equilibrium with the lattice. Noise measurements performed on the same organic solar cells in the dark confirm that without illumination, charges are in equilibrium with the lattice and give an electronic temperature ~300 K, cf. Fig. 3a. The upswing in noise spectral density towards higher frequencies is due to the capacitive contribution to the conductance that is relatively large in absence of photo-generated charges (Supplementary Text S3). We also ensured that the signals for the organic devices were not due to heating of the lattice by the illumination (Fig. S2). Similar measurements of noise with varying bias voltages were also performed on P3HT:PCBM solar cell and shown in Fig. S9c, d.

Converting the raw noise data from Fig. 2c,d to effective temperatures using Eq. (2) requires knowledge of the Fano factor, which accounts for correlations in the transport in the device and reduces the actual shot noise.[27] Unfortunately, no previous work has been done to determine the Fano factor in a three-dimensional system of the type at hand.[28] Using numerically exact kinetic Monte Carlo simulations, we estimate the Fano factor to be between 0.1 and 1 and recognize that it depends on temperature and electric field, as further discussed in Supplementary Text S4. Hence, a Fano factor determined in the dark or at short-circuit conditions would not be applicable at open-circuit. After subtracting the shot noise and cross-talk contributions, we measure an electronic temperature between 450 – 650 K for the two organic systems. The charge carrier temperatures in silicon were found to be ~305 K. In all cases, the lattice temperature was maintained at room temperature, i.e. 295 K. As such, Fig. 3b provides an upper limit for the electronic temperature at $F = 1$, while $F = 0.1$ provides a lower limit; the (unlikely) scenario in which $F < 0.1$ would not significantly decrease the temperatures further, as discussed in Supplementary Text S4. Having established that charge carrier populations in operational OPV devices have, in stark contrast to silicon, a temperature that lies significantly above that of the lattice does, in itself, not imply that OPV devices operate as HCSC in the way proposed by Ross and Nozik. To demonstrate that, we will use kinetic Monte Carlo (kMC) simulations to first show that, indeed, the enhanced electronic temperatures are due to the 'global' thermalization of photo-generated charges in a disorder-broadened DOS and, subsequently, that the so-called transport energy $\varepsilon_{tr}$ takes the role of the energy filter, as schematically shown in Fig. 4c.

The kMC method is an established way to simulate the extremely complex reality of large numbers of excitons and charges moving and interacting in the active layer of a macroscopic operational device. It does so by assigning probabilities to all possible events in a simulation box of finite size, here typically 30×30×55 sites, each with a random energy drawn from the Gaussian DOS. Using the calculated rates as weighting factors for each possible event (exciton generation and recombination, charge hopping, injection and extraction), a single event is randomly chosen and executed, after which the procedure is repeated. The method was previously used to reproduce a number of experiments on OPV devices, including full IV curves of PM6:Y6 bulk heterojunctions (BHJ) at different temperatures, ternary BHJ with and without composition gradients and non-equilibrium effects on recombination;[13,25,29,30] further details are given in Supplementary Text S6.

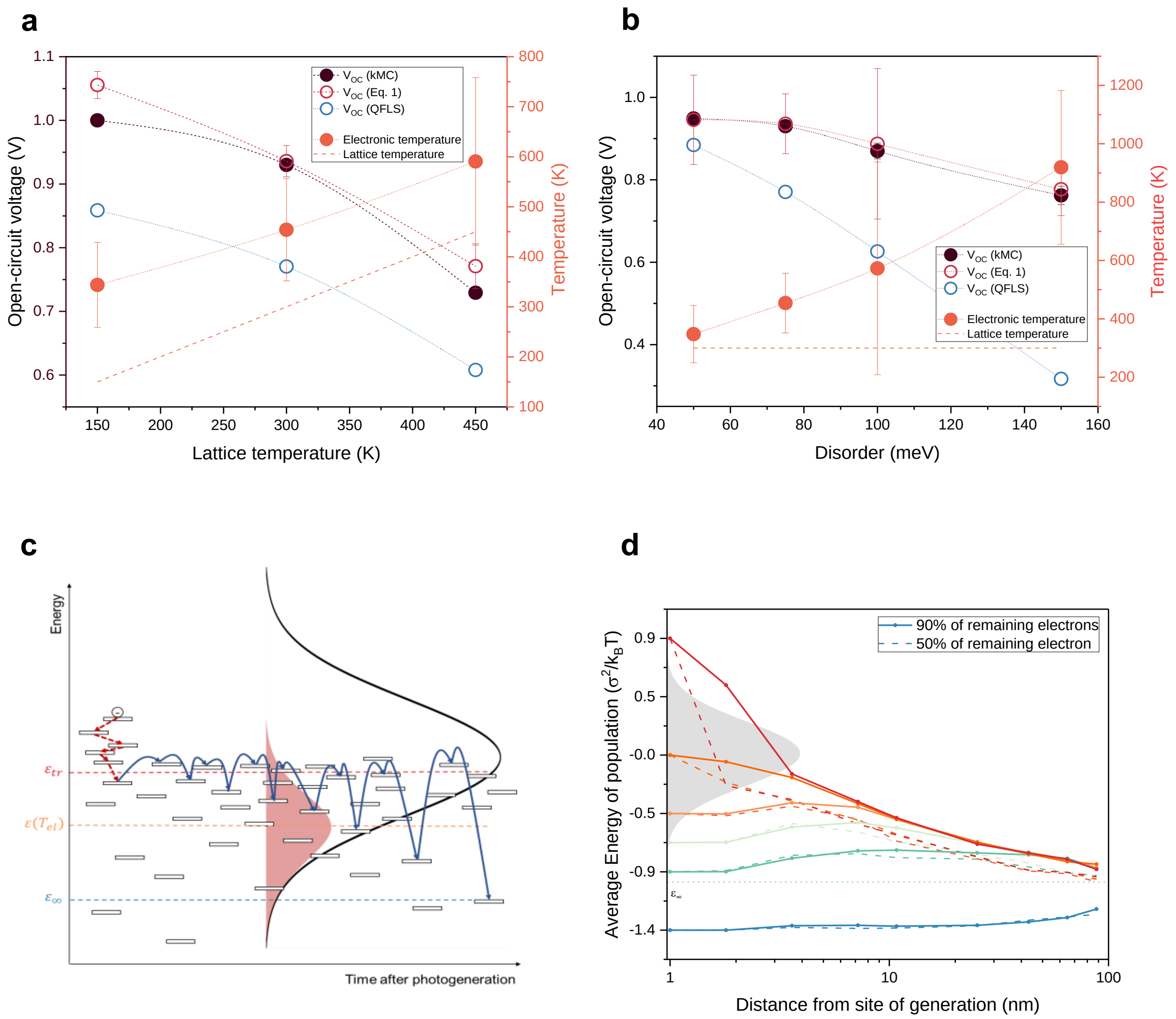

**Fig. 4 Implications of electronic temperatures higher than the lattice temperature and role of excess photon energy.** (**a**)&(**b**) Calculated open-circuit voltage for different lattice temperatures for $\sigma_{DOS} = 75$ meV (**a**) and different disorders for $T_{latt} = 300$ K (**b**) for the near-equilibrium model, kMC and the hot-carrier model. The open red circles indicate the open-circuit voltage values from Eq. 1, using the quasi-Fermi level splitting (open blue circles) and electronic temperature $T_{el}$ (orange circles) as input. The orange dashed lines show the lattice temperature. (**c**) Schematics of charge transport in disordered organic semiconductor. (**d**) Average energy of the carriers remaining after 10% (solid lines) and 50% (dashed lines) of the charges are extracted as a function of extraction distance for different excitation energies.

Building on our previous work, in which we used kMC to simulate IV-curves of OPV devices,[29] we extended the kMC model to simultaneously calculate current noise. Subsequently, a similar analysis as for the experiments was used to extract the electronic temperature at open-circuit conditions (Supplementary Note 7). The results are shown by black circles in Fig. 4a for different lattice temperatures and Fig. 4b for different energetic disorders, measured by the width $\sigma_{DOS}$ of the Gaussian density of states. In contrast to the experiments, the kMC simulations allow one to compare the noise temperature to the actual electronic temperature as determined by fitting $\sigma_{DOS}^2/k_BT_{el}$ to the mean energy of the photo-generated charge carriers, cf. Supplementary Text S8. Fig. S12 shows that the electronic temperature is consistent with the noise temperature of photo-generated charges.

As intuitively expected, Fig. 4a shows that the closer the system is to a band-like model, that is, the smaller $\sigma_{DOS}$, the closer the electronic temperature is to the lattice. However, even for the smallest disorder considered, 50 meV, a finite difference remains. Since disorder values reported for OPV materials typically fall in the range of 60-90 meV, the results presented here should apply to the vast majority of OPV devices.[21] This is also the reason why the two model systems that we used, P3HT:PCBM and PM6:Y6, give rise to similar electronic temperatures despite having very different device performance. The reason for the trends in Fig. 4a,b is that with increasing disorder or decreasing temperature, charges need exponentially longer to relax to global equilibrium as defined by the equilibrium energy.[25,26] The slowdown can, in turn, be understood via the concept of a 'transport energy', as illustrated in Fig. 4c. The transport energy $\varepsilon_{tr}$ is easily defined as the most likely energy that charges hop to in order to contribute to the charge transport. It was shown by Baranovskii et al. that for a broad class of strongly energy-dependent DOS, including, but not limited to Gaussians, the width of the transport path (in energy space) is narrow and that its center does not depend on the initial energy of the hopping charge.[31] Hence, by the definition of the transport energy, charges that have partially thermalized to an energy $\varepsilon$ ($\varepsilon_{tr} > \varepsilon > \varepsilon_{eq}$) need to be thermally excited to the transport energy to become mobile and have the possibility to find a lower energy. The associated waiting time scales with $\exp\left((\varepsilon_{tr} - \varepsilon)/k_B T\right)$, explaining the slowdown of the thermalization process with time and increasing disorder, see Fig. 4c and Fig. S3.

In the context of organic solar cells, the importance of the transport energy is that photo-generated charges are extracted at energies close to $\varepsilon_{tr}$, thus acting as an energy filter. In contrast to the Ross/Nozik idea, Fig. 1, this filter does not only sit at the contacts but is present throughout the device and is the result of the peculiarities of hopping transport in energetically disordered media. We used our kMC simulations to confirm the above and to determine the position of the transport energy $\varepsilon_{tr} \approx -\sigma_{DOS}^2/2k_BT$, relative to the center of the DOS, see Supplementary Text S9. Accordingly, $\Delta E_{use}$ in Eq. 1 is taken as the difference between the electron and hole transport energies, $\Delta E_{use} = \varepsilon_{tr}^e - \varepsilon_{tr}^h$. The parameter $\Delta\mu$ in Eq. (1) is the quasi-Fermi level splitting in the device, as seen from the perspective of an observer at $T_{el}$ and is determined by the difference in Fermi energies of electrons and holes, i.e. $\Delta\mu = \varepsilon_F^e - \varepsilon_F^h$. This is extracted directly from the kMC simulations (blue circles in Fig. 4a,b). Thus, with $T_{el}$ determined from noise simulations (orange dots in Fig. 4a,b), along with $\Delta E_{use}$ and $\Delta\mu$, we can compute the non-equilibrium $V_{OC}$ using Eq. 1 (open red circles), which coincides with the $V_{OC}$ as read from the IV-curves obtained from the kMC simulations (filled black circles). The fact that the $V_{OC}$ values from Eq. (1) almost exactly coincide with the $V_{OC}$ values from the kMC model is the main result of this Letter. It shows that 'common' organic solar cells are hot carrier solar cells.

Realizing that the difference between the transport energy and the Fermi energy over the electronic temperature is nothing but the entropy carried by a moving charge or, equivalently, the Seebeck coefficient,[32,33] allows to rewrite Eq. (1) as

$$eV_{OC} = \Delta\mu + 2S(T_{el} - T_{latt}), \qquad (3)$$

where we assumed the Seebeck coefficient $S = (\varepsilon_F - \varepsilon_{tr})/T_{el}$ to be equal for electrons and holes; otherwise, the factor 2 in Eq. (3) would be replaced by a sum over two terms.[34–36] Fig. S16 shows that the non-equilibrium $V_{OC}$ values calculated from Eq. (3) match the kMC values well. Eq. (3) has two important implications. First, the relatively large Seebeck coefficients in organic materials at low charge carrier densities, $S{\sim}800$ μV/K for typical OPV (Fig. S4), explain why electronic temperatures exceeding the lattice temperature by ~100 K lead to substantial enhancements in $V_{OC}$ of around 0.1 – 0.2 V, as found in Fig. 4a,b. Second, and more importantly, Eq. (3) has a transparent physical meaning in that the open-circuit voltage equals the near-equilibrium Fermi level splitting plus the electron and hole Seebeck voltages developing between the hot carrier populations in the device and the cold lattice. Therefore, any enhanced electronic temperature measured on an OPV device implies that it operates as an HCSC.

Hot-carrier contributions notwithstanding, state-of-the-art OPV devices, having PCE just below 20%, do so far not beat the Shockley-Queisser limit for a single, disorder-free absorber with a rectangular absorption onset (PCE 30%).[22] The reason is that disorder constitutes a loss channel, even if it is mitigated by hot-carrier effects. Still, the question arises whether these insights can be used to make more effective OPV. Intuitively, one might imagine that optically exciting the system with more energetic photons, producing electron-hole pairs with higher excess energies, or making the active layer thinner, giving charges less time to equilibrate, would increase the electronic temperature, and in turn, the open-circuit voltage. To this end, we calculated the average energy of photo-generated charges as a function of the distance between the generation and extraction point for different excitation energies. By plotting the mean energy after 10% (solid lines in Fig. 4d) and 50% (dashed lines) of the charges are extracted, we obtain a measure of the temperature of the remaining populations; details of this calculation are given in Supplementary Text S13.

The data in Fig. 4d confirms the previous result from Melianas et al. that any excess energy above the center of the DOS is lost within a few nm.[14] Excitations below the DOS center, but above the equilibrium energy, are longer lived, with transients converging after several tens of nm. Hence, one might expect minor $V_{OC}$ increases for device thicknesses $L$ below twice this distance (since the mean extraction distance is $\sim L/2$). Unfortunately, at these thicknesses, $V_{OC}$ actually drops due to incomplete absorption and charges diffusing to the wrong contact.[13] Nevertheless, Fig. 4d suggests that a rather efficient OPV device can be made by exploiting the fact that high-energy excitations actually do not significantly contribute to the enhanced electronic temperature because they quickly lose their excess energy. Hence, exciting the system with low-excess energy photons that predominantly excite electrons and holes in the lower half of the DOS, will hardly change the electronic temperature, and hence the $V_{OC}$ enhancement. For this to work, one needs a system where the energies of the first excited singlet state ($S_1$) and the charge transfer (CT) state are similar, which is the case for low driving force systems like PM6:Y6.[37,38] A simple estimate suggests that for a narrow excitation band around the $S_1$ absorption maximum at ~1.4 eV, a PCE of around 40% should be possible. Explicit kMC simulations confirm this simple estimate and, as discussed in Supplementary Text S14, show that narrow-band illumination around the transport energy further improves the device performance; only for even lower excitation energies, the lack of excess energy reduces both free charge generation and $V_{OC}$. While this is not immediately applicable to single-junction devices harvesting white (sun)light, it does offer new perspectives for organic multi-junction solar cells or applications where more narrow-band light is harvested, including smart windows or indoor PV.

## Conclusions

Summarizing, we measured the temperature of the charge carrier populations in operational organic solar cells by noise spectroscopy. The experiments prove that, for two representative model systems, the electronic temperature under illumination is almost twice that of the lattice, while the electronic temperature in an operational commercial silicon solar cell is equal to that of the lattice. Using kinetic Monte Carlo simulations, we reproduce the high noise temperature and confirm its relation to an enhanced electronic temperature. Building on the established theory for disordered semiconductors, we show that the hotness of the electron and hole populations is due to the slow thermalization in a broadened density of localized states, requiring increasingly difficult re-excitation to a relatively narrow transport energy, which thereby takes the role of an energy filter. With that, the charge and energy transport in an operational organic solar cell can be one-on-one mapped on the hot carrier solar cell concept by Ross and Nozik. Taking the electronic temperature and extraction energies as input, we quantitatively reproduce the enhancement of the open-circuit voltage over its equilibrium value. These findings exemplify that typical organic solar cells, including the current state-of-the-art, are hot carrier solar cells. While the ultimate goal of a HCSC is to surpass the detailed balance limit, this work is a step towards this goal in demonstrating the importance of hot-carrier effects in OSC, and in particular their contributions to $V_{OC}$.

Since organic solar cells have so far almost uniquely been optimized on the basis of loss analyses assuming near-equilibrium,[39] the notion that charge carrier populations are actually hot, and that this hotness contributes to $V_{OC}$, not only necessitates a rethink but also greatly widens the scope of strategies to further improve these devices.[40,41] First, and most straightforward, our work suggests noise spectroscopy as a probe for energetic disorder, the reduction of which should, hot carrier effects notwithstanding, lead to higher $V_{OC}$ as seen in Fig. 4b. Second, and more interesting, the connection between the Seebeck effect and $V_{OC}$, Eq. (3), offers a novel way towards high-performance OPV devices. Techniques from the seemingly unrelated field of organic thermoelectrics, such as DOS engineering,[42,43] and morphology controlling[44] could be used to enhance $S$ of the blend of organic materials that form the active layer of the OPV device.

Beyond organic PV, we expect our findings to inspire new avenues towards high-efficiency harvesting of solar energy.

## Author contributions

P.V. performed the noise experiments; P.V. and C.G. fabricated samples and performed sample characterization; P.V. and C.T. performed the numerical simulations; P.V., C.T. and M.K. wrote the manuscript; M.K. conceived the idea and supervised the project. All authors contributed to the data analysis and conceptualization.

## Conflicts of interest

Authors declare that they have no competing interests.

## Data availability

The data supporting this article have been included as part of the manuscript and its Supplementary Information.

## Acknowledgements

This work has been funded by the German Research Foundation under Germany`s Excellence Strategy (2082/1 – 390761711, C.T.). M.K. thanks the Carl Zeiss Foundation for financial support. We are grateful to Olle Inganäs and Dieter Neher for discussions and critical reading of the manuscript and Sebastian Klein for noise floor characterization.

# Supplementary Materials for

## Hot Carrier Organic Solar Cells

Priya Viji, Constantin Tormann, Clemens Göhler, Martijn Kemerink*

*Corresponding author: martijn.kemerink@uni-heidelberg.de

**The PDF file includes:**

Materials and Methods

Figs. S1 to S17

Supplementary Texts S1-S14

Table S1

## Materials and Methods

### Device Fabrication

The silicon PV device used in this work was purchased from ThorLabs and is used as is. For the preparation of OSC, pre-patterned ITO-coated glass substrates were treated with oxygen plasma for 5 min, after which PEDOT:PSS (AI 4083; purchased from Heraeus) was spin-coated at 3000 rpm for 30 s. PM6:Y6 was dissolved in 1:1.2 (w/w)ratio in chloroform at 15 mg/mL concentration with 0.5% chloronapthalene as an additive. PFN-Br was chosen as the electron transport layer and was spin-coated from a 0.5 mg/mL solution in methanol. Finally, 75 nm of silver was evaporated as top contact under high vacuum conditions using a thermal evaporator. P3HT:PCBM was dissolved in 1:1 (w/w) ratio at 20 mg/mL concentration in chlorobenzene. 5 nm of calcium, followed by 75 nm of aluminum, was evaporated to complete the solar cell. The final area of the devices is 4 $mm^2$.

### Device Characterization

All devices were encapsulated in a glovebox ($O_2$ and $H_2O$ < 1 ppm). The current density-voltage (jV) characteristics were recorded in a glovebox from negative to positive bias in steps of 10 mV using a Keithley 2636B source meter under AM1.5G solar simulator from Abet Technology with an ABA rating. See Fig. S1.

### Sensitive EQE Measurement

Monochromatic light from a PerkinElmer Spectrophotometer Lambda 1050+ is modulated via an optical chopper at 263 Hz falls on the device under test (DUT). The current generated by the DUT at short-circuit conditions is amplified using a low-noise transimpedance amplifier from Femto (DLPCA-200) before being fed to a lock-in amplifier from Zurich Instruments (MFLI). The setup was calibrated against reference silicon, germanium, and indium-gallium-arsenide photodiodes. See Fig. S1.

### Noise Measurement

The DUT is kept in a Linkam HFS600E miniature cryostat which acts as both a Faraday cage as well as helps to keep the temperature constant under constant illumination. Measurements are performed at constant illumination of 0.03 Sun for P3HT:PCBM and PM6:Y6. A halogen lamp is used as the light source. The low illumination level was used in order to have a reasonable measurement time (see Supplementary Text S2). Stanford Research Systems transimpedance amplifiers converted the current from DUT to voltage while also applying a constant bias to the DUT. Two time-synchronized lock-in amplifiers from Zurich Instruments collected the time domain data. More details are given in Supplementary Text S1 and S2.

## Supplementary Figures

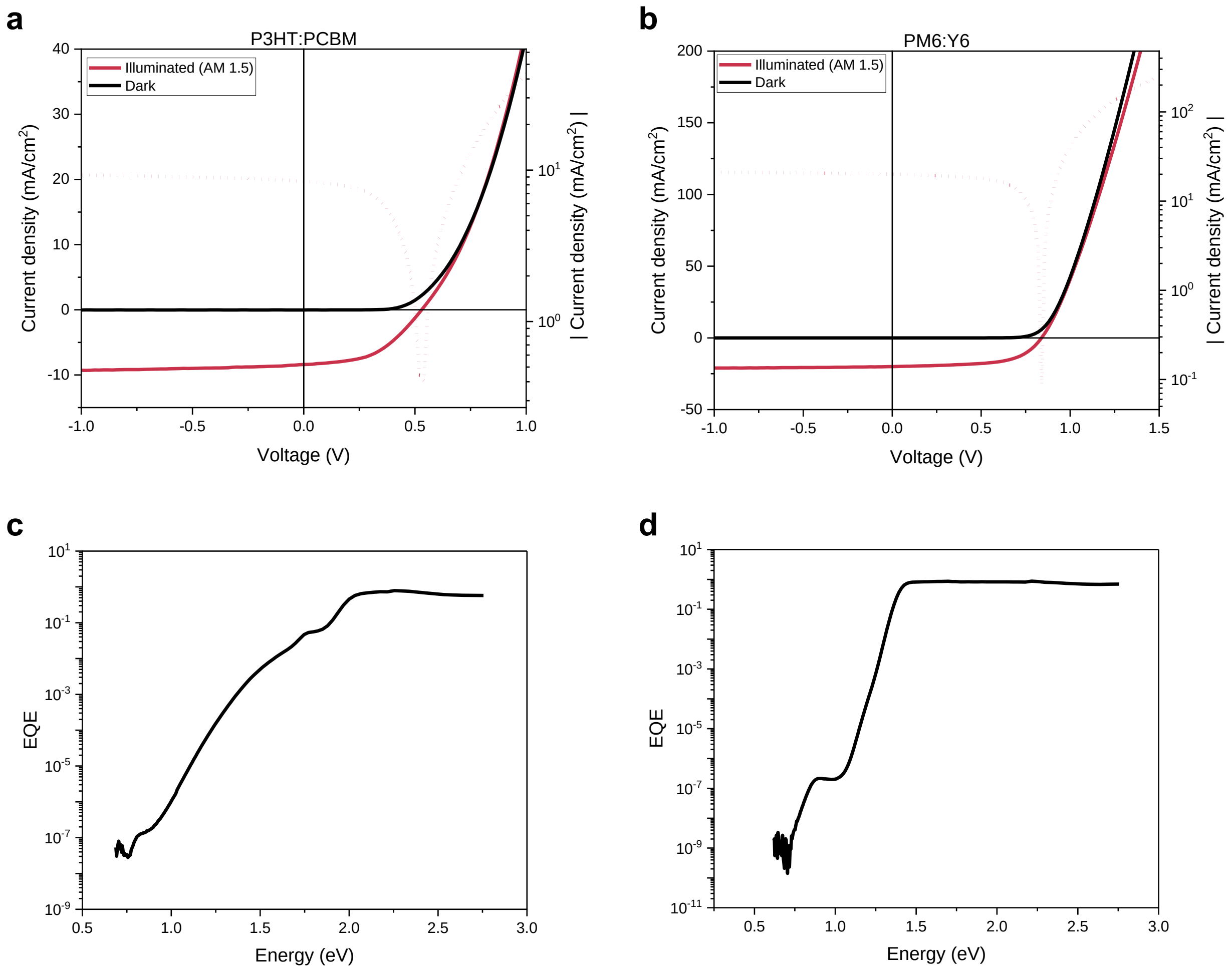

**Fig. S1 j-V characteristics and EQE of OSC.**
Current density versus voltage characteristics of (**a**) P3HT:PCBM and (**b**) PM6:Y6 solar cells under 1 sun AM 1.5G illumination at room temperature. External quantum efficiency (EQE) spectra of (**c**) P3HT:PCBM and (**d**) PM6:Y6 at room temperature.

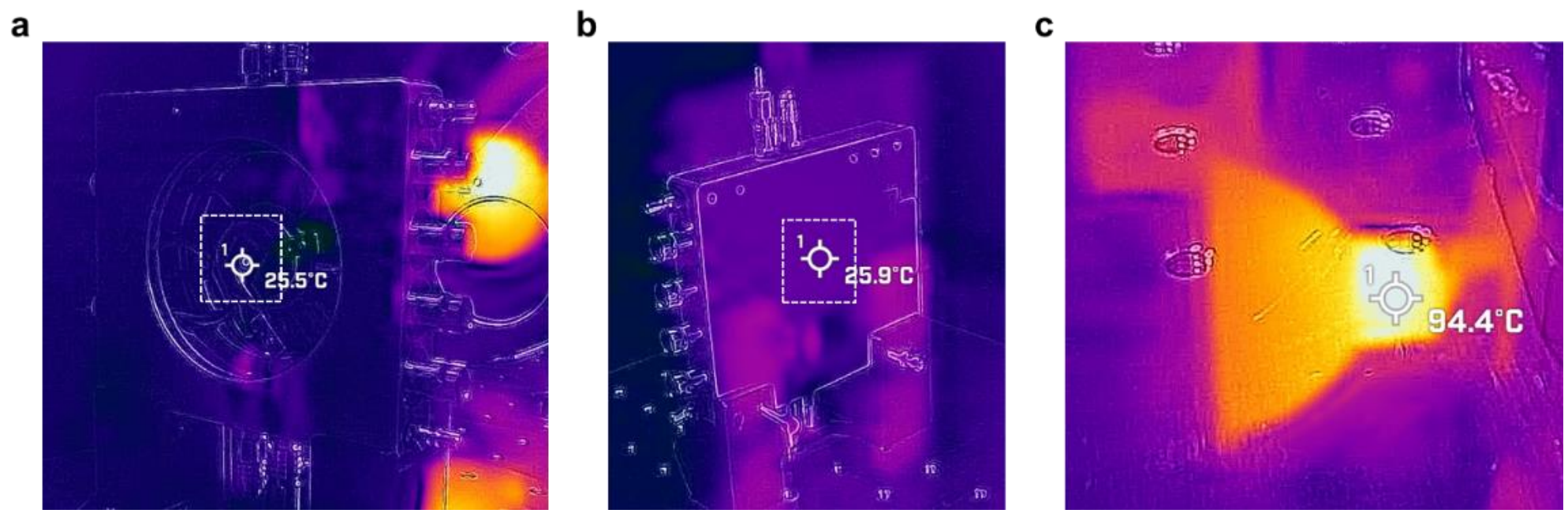

**Fig. S2 Thermographic images of OSC (without contacts) under illumination.**
In order to rule out heating effects due to illumination with a halogen lamp, thermographic images were taken of the setup after ~3 hours of illumination. The sample is mounted in a metallic stage and illuminated through a small hole in the bottom. (**a**) Thermal image of the DUT on the back side of illumination. (**b**) Thermal image on the side of illumination. The DUT is mounted on the opposite side. (**c**) The temperature at the base of the halogen lamp after 3 hours of operation.

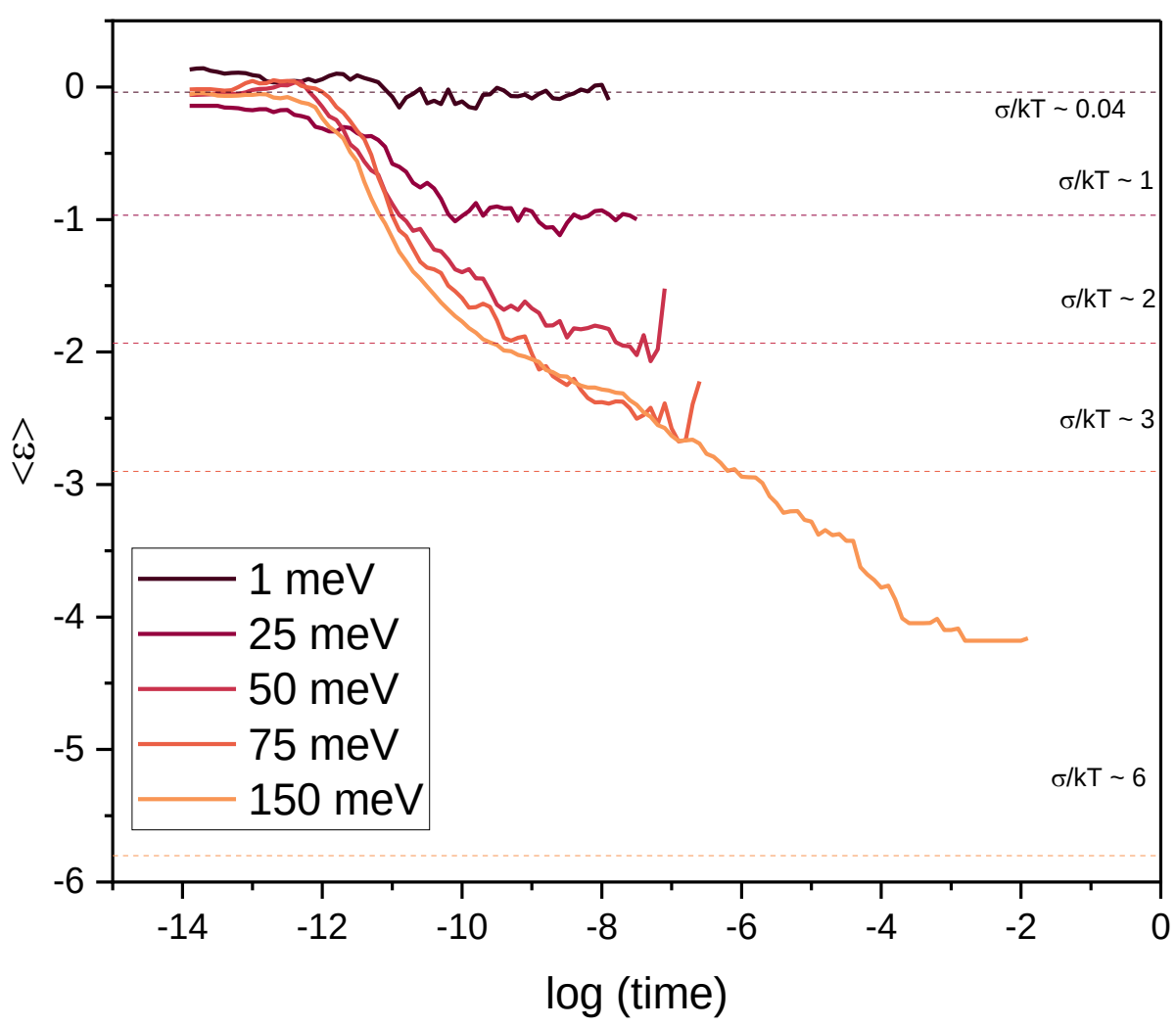

**Fig. S3 Calculated temporal evolution of the energy of photo-generated charges in a solar cell with extracting contacts after an initial light pulse.**
<ε> is the average energy of photo-generated charges created via a light pulse in units of $\sigma_{DOS}/k_B T$. For a low-disorder system, $\sigma \leq 50$ meV, the charges thermalize before they are extracted, while in higher disordered systems, charges are extracted before fully thermalizing and hence, their electronic temperatures are higher. The last data point for each disorder represents the last (slowest) charge that left the simulated device. Attempt to hop frequency $\vartheta_0 = 1\times10^{11}$ $s^{-1}$ and $T = 300$ K.

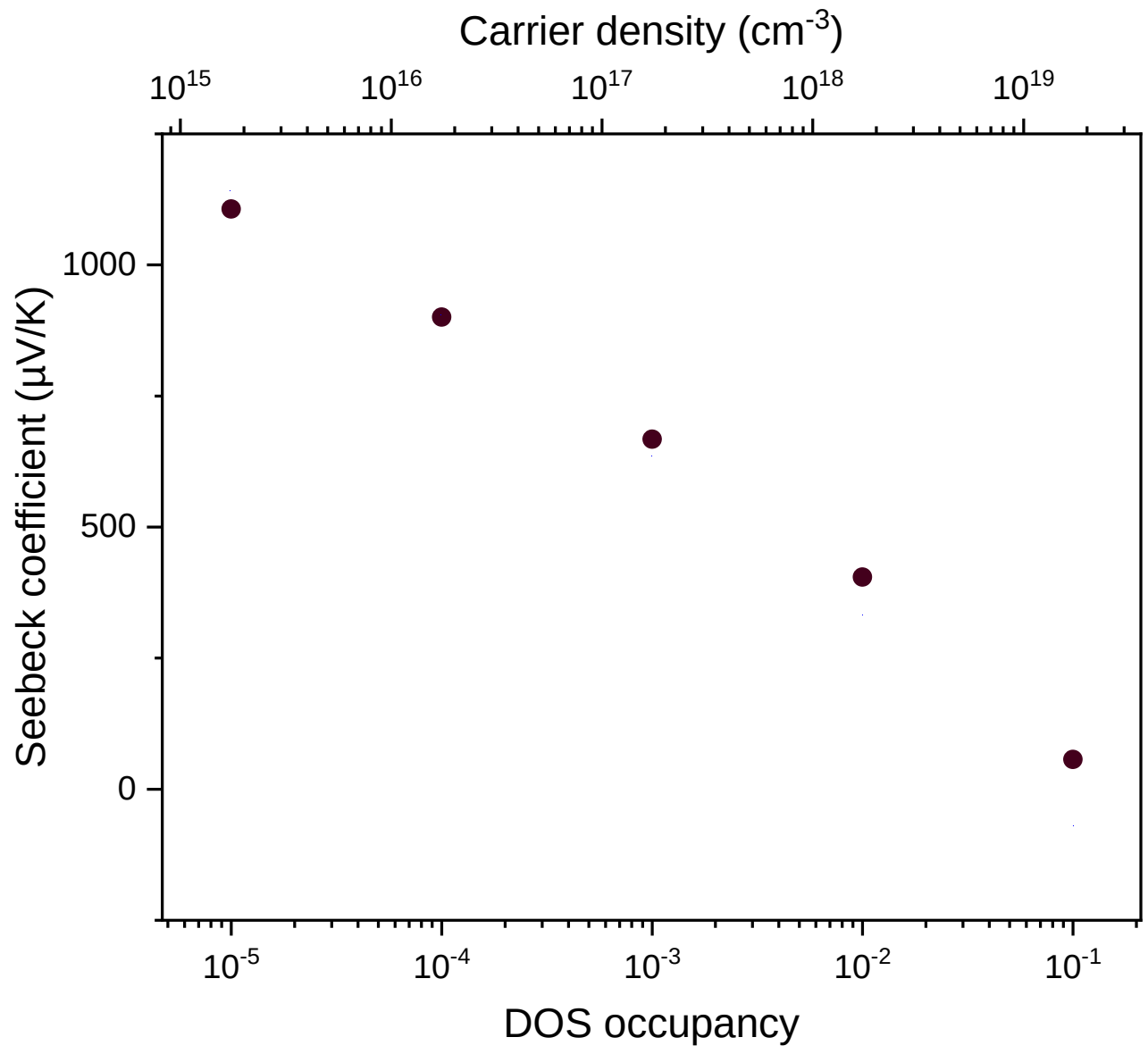

**Fig. S4 Variation of the Seebeck coefficient as a function of carrier concentration.**
Seebeck coefficients calculated in kMC using $S = (\varepsilon_F - \varepsilon_{tr})/T$, where $\varepsilon_F$ is the Fermi energy and $\varepsilon_{tr}$ is the transport energy. The simulations were performed at $1 \times 10^7$ V/m external electric field and a lattice temperature of 300 K.

## Supplementary Text

## S1: Cross-Correlation Noise Spectroscopy

The cross-correlation noise spectroscopy methodology to perform Johnson thermometry was based on the works of Sampietro et al.[1] The current fluctuations from the DUT are fed, using low-noise cables from Femto Messtechnik GmbH, to two identical but independent transimpedance amplifiers by Stanford Research Systems (SRS 570) with variable gain. The amplifiers are also used to bias the DUT; thus, an external biasing system that would introduce additional noise was not required. The amplified voltage signals are independent and are connected to lock-in amplifiers from Zurich Instruments (MFLI), which are time-synchronized to capture the signals simultaneously. This way, the additional, uncorrelated noise added by the independent transimpedance amplifiers can be removed from the measurement. The amplifiers run on batteries to reduce the 50 Hz hum and were kept floating to avoid ground loops. The time-domain signals were acquired with the MFLIs using a Hann window and accordingly corrected with a scaling factor of 1.5.[2] The calculation of the power spectral density (PSD) from the signals acquired by MFLIs is shown below.

Let the noise from DUT be represented by $n(t)$ and the noise introduced by the two transimpedance amplifiers be $a(t)$ and $b(t)$. By definition, $n(t)$, $a(t)$ and $b(t)$ are statistically independent. Thus, the signal observed at the end of channel one be $x(t)$ and channel two be $y(t)$ and is given by

$$x(t) = n(t) + a(t)$$

$$y(t) = n(t) + b(t)$$

Cross-correlation measures the similarity of two signals as a function of time/frequency. One can understand it as a flipped convolution. The PSD is calculated as

$$S(f) = \mathcal{F}\,[x(t)]^* \times \mathcal{F}[y(t)]$$

$$= (\mathcal{F}[n(t)] + \mathcal{F}[a(t)])^* \times (\mathcal{F}[n(t)] + \mathcal{F}[b(t)])$$

$$= |\mathcal{F}[n(t)]|^2 + \mathcal{F}\,[n(t)]^*\mathcal{F}[b(t)] + \mathcal{F}\,[a(t)]^*\mathcal{F}[n(t)] + \mathcal{F}\,[a(t)]^*\mathcal{F}[b(t)]$$

$$\mathbb{E}\left\{S_{yx}(f)\right\} = \mathbb{E}\{|\mathcal{F}[n(t)]|^2\} + \mathbb{E}\{\mathcal{F}[n(t)]^*\mathcal{F}[b(t)]\} + \mathbb{E}\{\mathcal{F}[a(t)]^*\mathcal{F}[n(t)]\} + \mathbb{E}\{\mathcal{F}[a(t)]^*\mathcal{F}[b(t)]\}$$

Here, $\mathcal{F}\,[x]$ represents the Fourier transform of the signal, $x$, $x^*$ is the complex conjugate of $x$, and $\mathbb{E}$ is the expectation value. Since $n(t)$, $a(t)$ and $b(t)$ are statistically independent,

$$\mathbb{E}\{\mathcal{F}[n(t)]^*\mathcal{F}[b(t)]\} = \mathbb{E}\{\mathcal{F}[a(t)]^*\mathcal{F}[n(t)]\} = \mathbb{E}\{\mathcal{F}[a(t)]^*\mathcal{F}[b(t)]\} = 0$$

Therefore, it follows that

$$\mathbb{E}\left\{S_{yx}(f)\right\} = S_n$$

When averaged over many repeated measurements, $N$, the averaged noise power is given by

$$S_{avg}(f) = \frac{\sum_{n=1}^{N} S_n(f)}{N} \qquad \text{(S1)}$$

Thus, a correlation spectrum results from processing two independent signals from the same DUT and taking advantage of uncorrelated noise properties of the input stages.

In this work, the gain of the transimpedance amplifier was set to $10$ μA/V, which gives a noise floor of $7.84 \times 10^{-26}$ A²/Hz for 100 correlation steps, while the lowest noise floor measured with this setup was $8.4 \times 10^{-30}$ A²/Hz using a gain of $100$ nA/V for 24000 correlation steps.

## S2: Noise Spectroscopy Setup

The DUT is housed within a Linkam HFS600E miniature cryostat under a nitrogen atmosphere, which acts as a Faraday cage and maintains a stable temperature under continuous illumination. The measurements are performed at a constant illumination of 0.03 Sun for the P3HT:PCBM, PM6:Y6 and Si PVs. Low illumination intensities are chosen to maintain the consistency in the gain setting on the transimpedance amplifier at 10 μA/V. Higher illumination levels would lead to increased currents, necessitating lower gain settings, which in turn would elevate the noise floor (Fig. S5). In order to measure the DUT's noise at 1 Sun and at a single voltage, we would need to increase the number of correlation steps to at least 20 000 repetitions, which will take about 30 hours until the measurements can be reliably differentiated from the noise floor. With the settings used, the measurement for each biasing voltage took 10 minutes. Importantly, higher illumination intensities do not affect the thermalization kinetics and hence do not affect the electronic temperature.[3,4]

We used a halogen lamp as our light source, the most noiseless light source we tested. Although it is hard to quantify noise levels of illumination sources when these are below the noise level of the measurement device, the fact that the measured noise temperature of the illuminated silicon PV device coincides with the ambient temperature implies that the noise from the halogen lamp does not notably affect the measurements presented herein. Additionally, the shot noise limit of all PVs used in the work was reached, confirming that the light intensity fluctuations from the halogen lamp are lower than the noise of interest.[5] The setup was mounted on an aluminium breadboard to which the optical components were securely attached. In order to avoid benchtop vibration, the breadboard was mounted on Sorbothane feet.

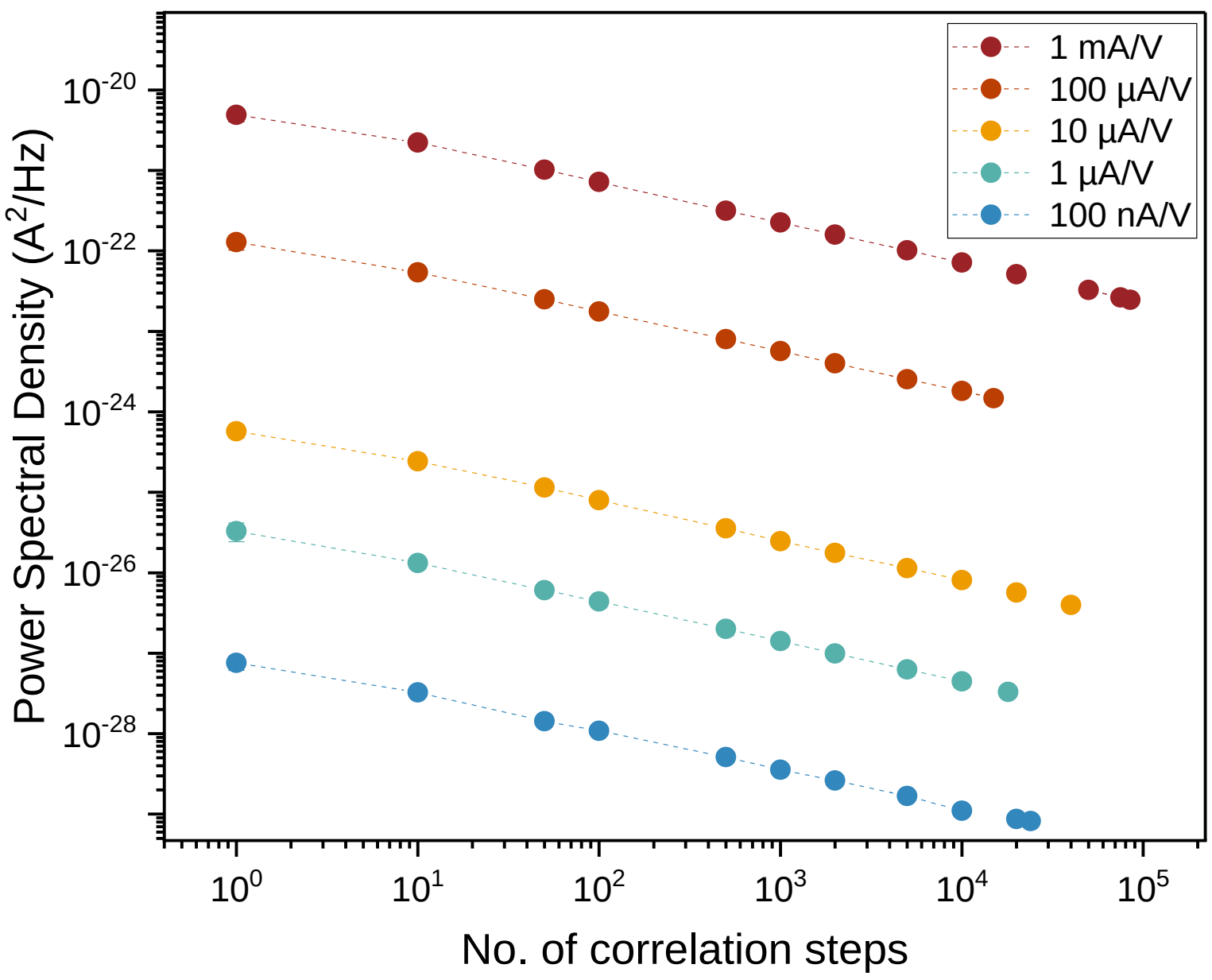

**Fig. S5 Validation and noise floor of the setup.**
The noise floor for the home-built noise spectroscopy setup for different gains of transimpedance amplifiers as a function of the number of correlation steps.

## S3: Determining the Temperature of Solar Cells in the Dark

Throughout optimizing the noise measurement setup, we checked the temperature of simple carbon resistors and organic semiconductors like P3HT, with and without doping and always found, within experimental uncertainty, the electronic temperature to be equal to the lattice temperature (see Fig. S6a). In order to assess whether the organic solar cell in the dark is also at lattice temperature, we measured noise at various voltages. Due to the dominating shot noise after the 'knee-point', we measured at or around $V = 0$ V. The impedance of the solar cell is dominated by the capacitive impedance and, therefore, leads to a frequency dependence; hence, extracting noise from this shape is quite tricky. We performed an impedance spectroscopy of an OSC with 100 mV of AC voltage so that we were well below the knee point. We perform current cross-correlations so the measurement remains comparable to the other results. Since the current in the system is very low, we use a gain of 100 nA/V, and have to perform about 2000 repetitions to sufficiently lower the noise floor to differentiate between the noise floor and the noise of the DUT. Fitting the measured data with various temperatures revealed the electronic temperature to be close to the lattice temperature. As temperatures below lattice temperature are impossible, we can safely assume that the solar cell in the dark is not hot. The data shown in Fig. S6b was taken for a cell with lower shunt resistance than the OPV presented in the main text.

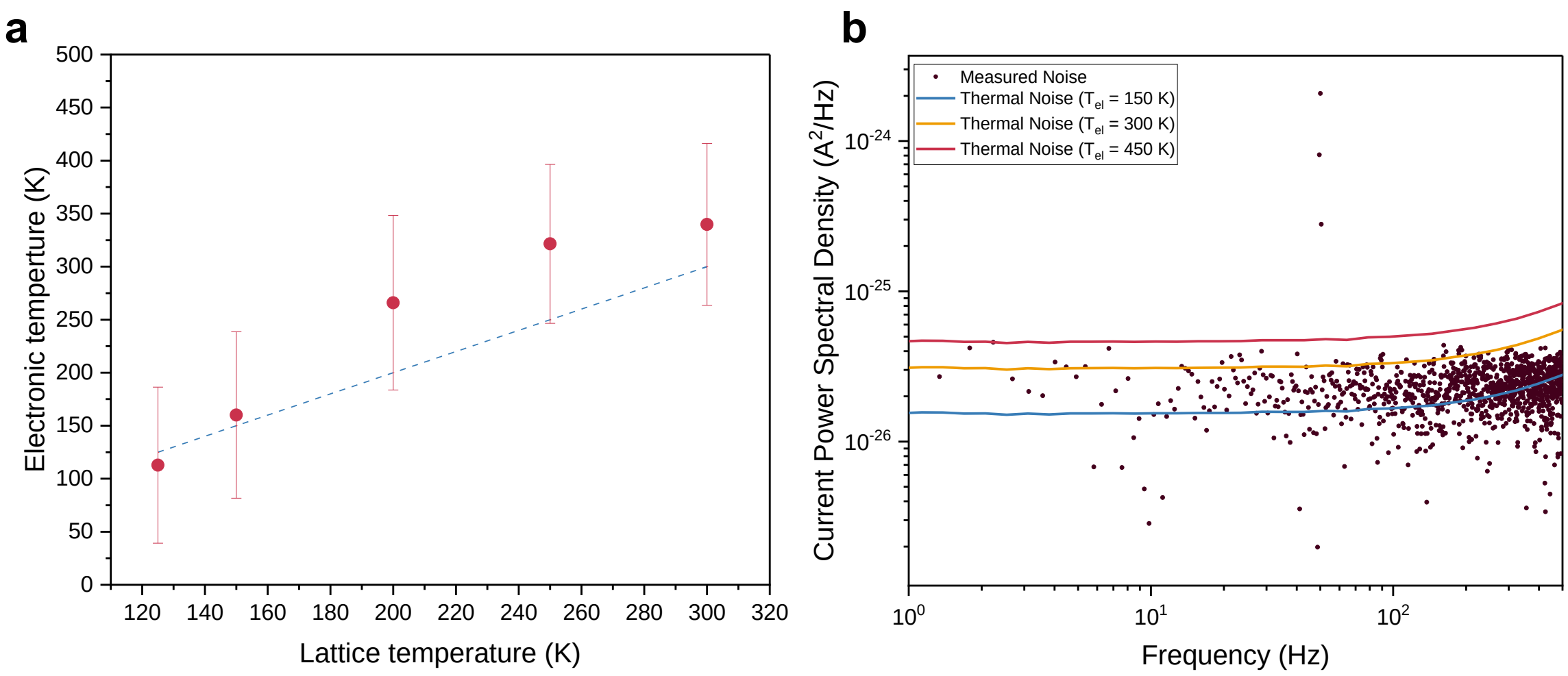

**Fig. S6 Electronic temperature of doped P3HT and noise spectra of OSC in the dark.**
(**a**) The electronic temperature of an in-plane P3HT thin film with 0.1 mg/mL doping of $F_4TCNQ$ was measured for varying lattice temperatures. The blue dashed line indicates the lattice temperature.
(**b**) The black dots are the measured noise of an organic solar cell in the dark, while the colored lines are the thermal noise calculated from $4k_BT_{el}/Z$ using the measured $Z$ from impedance spectroscopy and varying $T_{el}$.

## S4: Fano Factor

Walter Schottky discovered shot noise in vacuum tubes in 1918 while studying the emission of discrete electrons from the cathode.[6] Emission is a random, independent Poissonian process, and the noise was found to be proportional to twice the electric charge and the net current. Subsequent research revealed that correlations in the process can reduce the shot noise by the Fano factor, $F$

$$S_{non-Poissonian} = 2 \times q \times \bar{I} \times F \qquad \text{(S2)}$$

where $\bar{I}$ is the average current flowing through the device. Coulomb repulsion and Pauli's principle preventing double occupation are two possible effects leading to correlated processes.

In organic semiconductors, charge transport occurs via hopping conduction, where charges tunnel between localized sites. It is much simpler to visualize a 1-D system with N identical sites, represented by a resistor network with identical resistance $R$; the Fano factor would be 1/N. However, in reality, these resistors are different. If the transport is only limited by the most resistive hop (hard hop), then $F = 1$. However, for a real system, transport can be understood in terms of percolation theory, in which percolation clusters of various sizes and sites at the percolation threshold determine the hard hops. The diminution of $F$ can be related to the length of an average percolation cluster in the disordered material to the total length of the conductor.[7–9]

While there has been no experimental determination of the Fano factor in organic semiconductors, we looked for similar device structures like photovoltaics or LEDs. In Davenport et al.'s work,[10] the Fano factor in a perovskite solar cell at short-circuit conditions was determined to be sub-unity (~0.8). In the case of organic LEDs,[11] $F$ as function of biasing voltage was measured and dropped from 1 to 0.5.

From the mentioned studies, we conclude that the Fano factor varies with the biasing in the OSC because charges perceive a different energy landscape at short-circuit compared to open-circuit. To confirm our hypothesis, we performed kMC simulations of randomly distributed charges in a box at different electric fields (without contacts). Fig. S7 shows the resulting variation of the Fano factor.

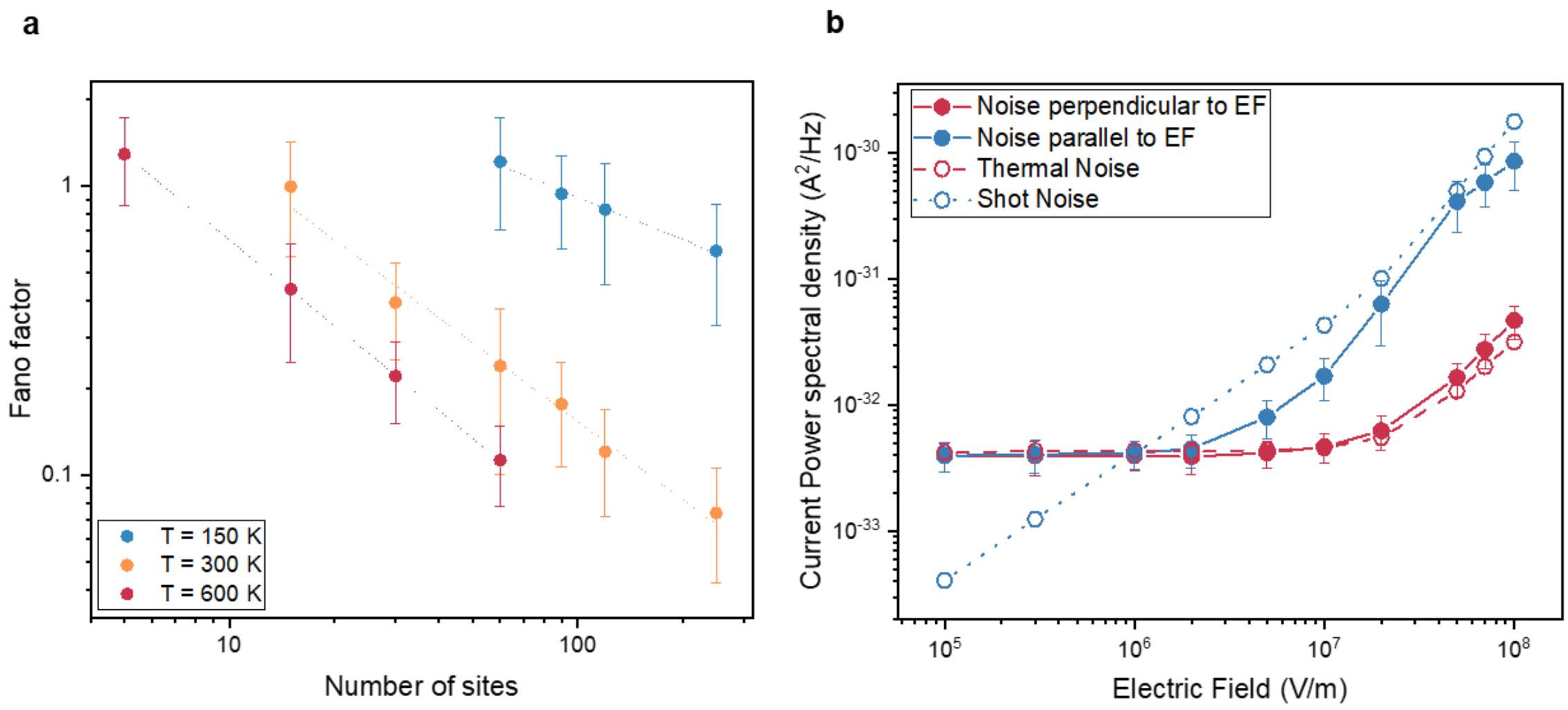

**Fig. S7 Variation of Fano factor.**
(**a**) Fano factor calculated from kMC simulations of randomly distributed charges in a box of varying length. The number of sites in the field direction was varied for different lattice temperatures while maintaining a constant electric field (EF) of $1 \times 10^7$ V/m. (**b**) Noise vs electric field calculated using kMC simulations. Note that at low fields, the thermal noise is dominant. At higher fields, shot noise (reduced by $F$) increases the overall noise in the field direction, while in the direction of zero current flow, thermal noise remains the only noise source. Thermal noise increases at very high fields ($>$ $5 \times 10^7$ V/m) due to higher effective temperatures stemming from field heating.[12]

In a 200 nm thick PM6:Y6 device used here, we have about 100 sites in the electric field-direction (assuming an inter-site distance of 1.8 nm). If all the hops were equivalent and equally hard, the number of hard hops in the device would be 100, leading to $F$ ~1/100, which is the lower limit of the Fano factor. Even in this exaggerated scenario, the electronic temperature does not deviate much compared to $F = 0.1$ and is within the error limits. Suppressing the shot noise any further with lower $F$ is irrelevant because the measured shot noise is already an order of magnitude lower than the thermal noise at open-circuit (see Fig. S8).

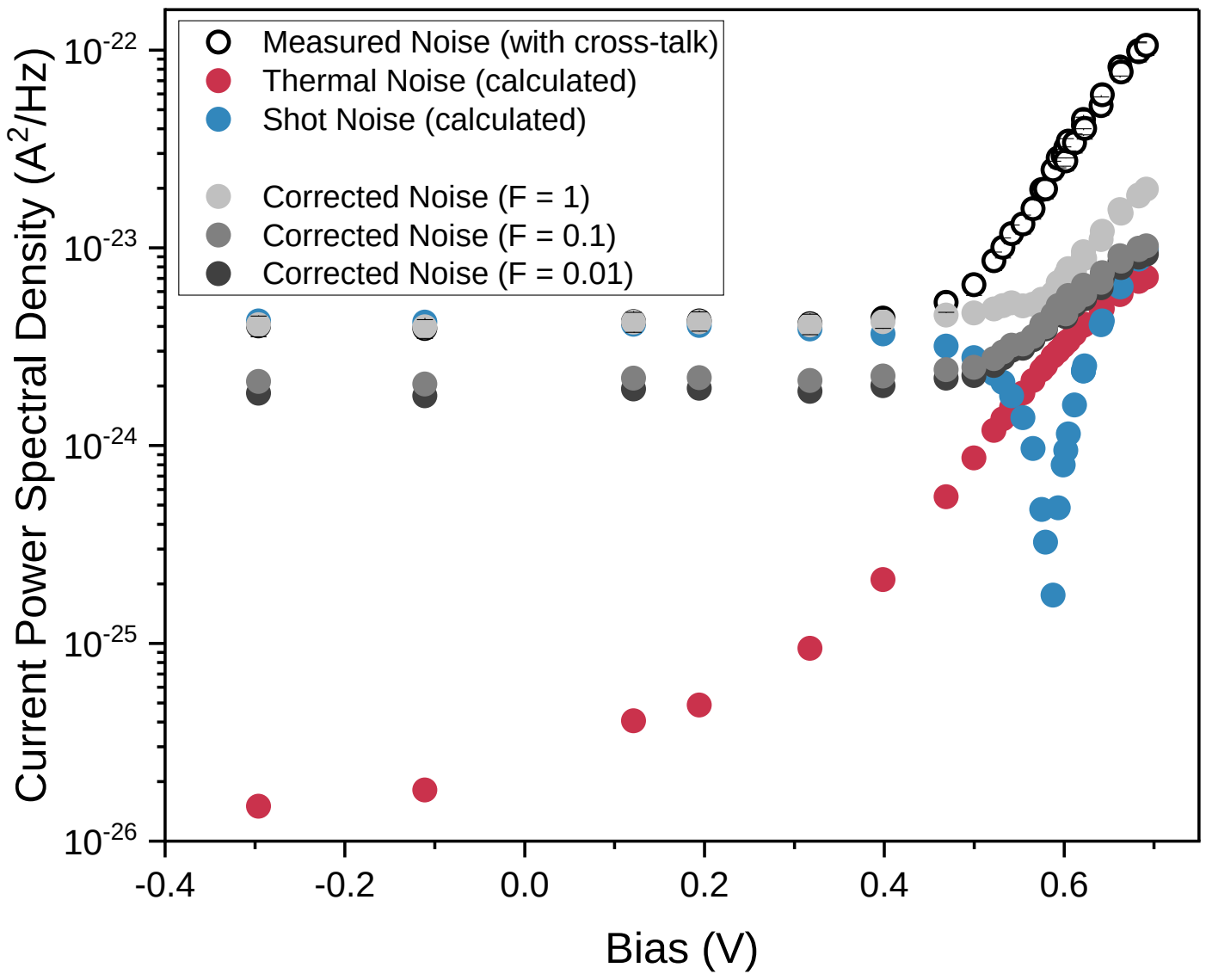

**Fig. S8 Estimation of the influence of the Fano factor.**
Noise vs bias voltage for PM6:Y6 under illumination for different Fano factors. Further discussion on cross-talk is given in Supplementary Text S5.

While there have been some studies investigating the Fano factor in silicon,[13] they were conducted under short-circuit conditions. Given that the exact composition of commercially bought silicon photodiode is unknown, we refrain from attempting to determine the Fano factor. Notably, a Fano factor of 0.44 was reported for a high-performance, multi-layered Si solar cell at short-circuit,[14] while we observe 0.95. Thus, we proceed with the same assumption as for OSCs, namely that $F$ lies between 0.1 and 1 for silicon. This gives us an $T_{el} = 295\ \mathrm{K}\ (F = 0.1)$ and $T_{el} = 302.5\ \mathrm{K}\ (F = 1)$ as seen in Fig. 3b.

## S5: Current Noise over Voltage Noise and Cross-Talk

Thermal noise in any device can be accessed by measuring current or voltage fluctuations in the DUT. Measuring current noise over voltage noise was preferred for this work due to the possibility of biasing the DUT directly using the transimpedance amplifier, thereby avoiding additional biasing circuitry. Additionally, the kMC simulation conducted in this study also evaluates the current noise, facilitating a direct comparison. However, it is worth noting that current cross-correlation also presents certain drawbacks, which will be discussed in this section.

Even for a two-channel noise spectroscopy system, there can be correlated spurious signals generated by the transimpedance amplifiers that limit the sensitivity of the setup. According to Eq. 2 in Ferrari et al.,[15] the cross-talk noise (or correlated noise) increases as the impedance of the DUT decreases. Given that the impedance of a solar cell varies significantly for different biasing conditions, both current and voltage noise measurements are inherently constrained. Performing current noise measurements works absolutely well in short-circuit conditions when the resistance of the device is high, and it enables accurate measurement of the shot noise of the DUT. However, the noise is exaggerated for measurements around the open-circuit due to increased cross-talk (Fig. S8).

To quantify the contribution of cross-talk to the measured noise at open-circuit, the devices are first measured in the dark. The underlying assumption is that noise due to cross-talk is independent of illumination and the type of device (OSC, carbon resistors, silicon PV), as it is only dependent on the DUT's resistance. Fig. S6b shows that the dark temperature agrees well with a 300 K thermal noise fit. For each specific voltage, it is now possible to quantify the excess noise due to cross-talk as we know the resistance, the current, the lattice temperature and the measured noise. We define the following cross-talk correction factor

$$f_{CT} = \frac{Measured\ Noise\ (dark) - F \times Shot\ Noise\ (dark)}{Thermal\ Noise\ (dark)}$$

where we can subtract the shot noise because it is not influenced by the cross-talk. Since we do not know the exact Fano factor, we use 0.1 and 1, which represent the lower and upper limits (see Supplementary Text S4), respectively, for the shot noise. Repeating the measurement for different voltages, we quantify the excess noise due to cross-talk as a function of resistance for each device, allowing us to correct the measurements conducted under illumination. The corrected noise for measurements performed under illumination is then given by

$$\begin{aligned} &Corrected\ Noise\ (illu.) \\ &= \frac{Measured\ Noise\ (illu.) - F^{*} \times Shot\ Noise\ (illu.)}{f_{CT}} + F^{*} \times Shot\ Noise(illu.) \end{aligned}$$

where the first term represents the corrected thermal noise. It should be noted that the Fano factor ($F^{*}$) under illumination can be different from the one in the dark. We, again, estimate an upper and a lower limit of 1 and 0.1. The corrected electronic temperate is finally calculated via

$$T_{el} = \frac{Corrected\ Noise\ (illu.) - F^{*} \times Shot\ Noise}{4k_B/R} \qquad (S3)$$

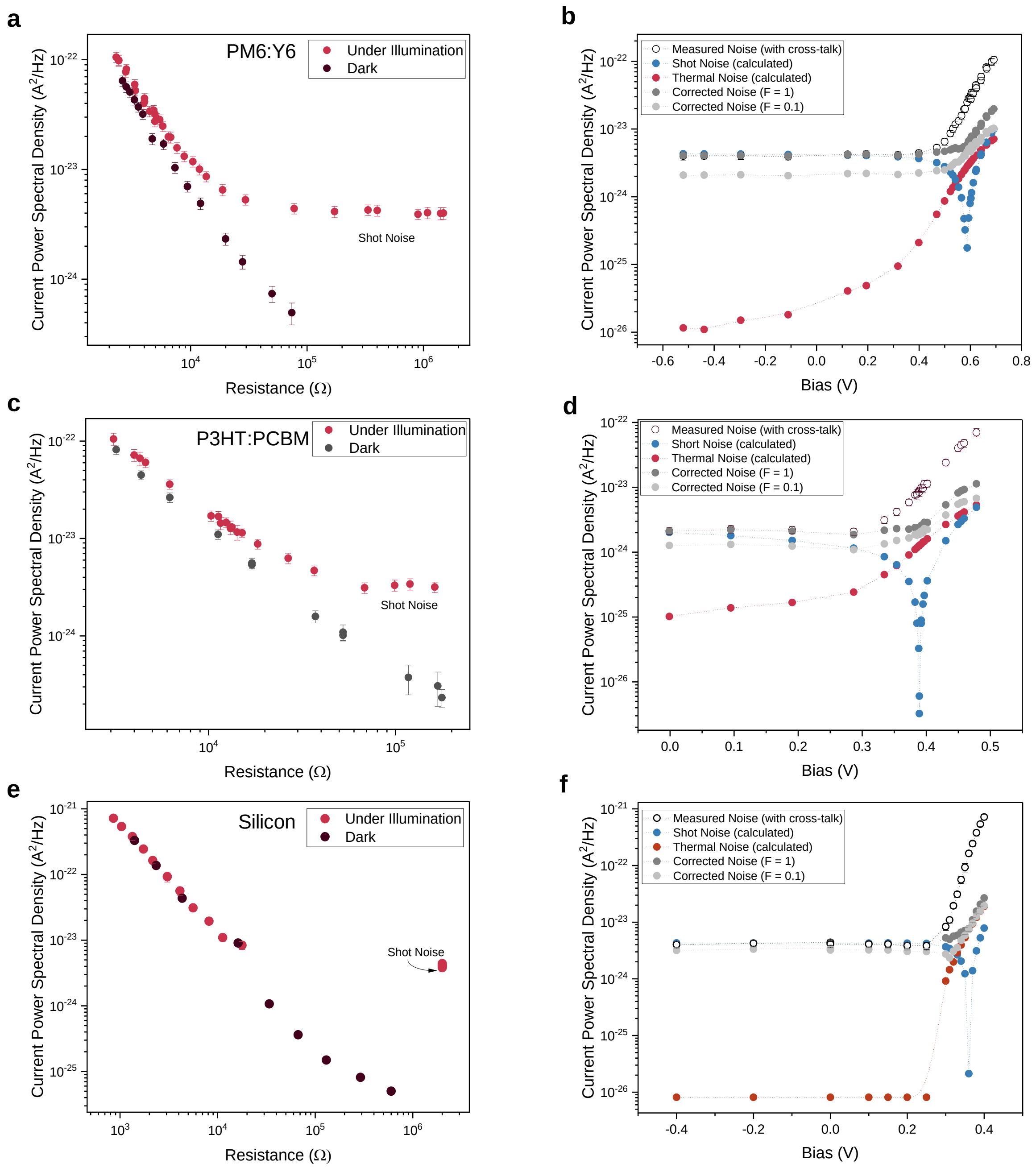

**Fig. S9 Cross-talk effect in the measurement system.**
(**a**), (**c**) & (**e**) Measured noise (before corrections) as a function of differential resistance of PM6:Y6, P3HT:PCBM and silicon, respectively. Note how the red and black circles deviate in (**a**) & (**c**), while in (**e**) they lie on top of each other, indicating the difference in temperature in the dark and under illumination for two kinds of OSCs, while they are almost equal for silicon. The plateau in the noise under illumination is the shot noise contribution. (**b**), (**d**) & (**f**) Noise vs bias voltage for PM6:Y6, P3HT:PCBM and silicon, respectively, with and without cross-talk correction and with $F = 0.1$ and 1.

## S6: Kinetic Monte Carlo Simulations

The kinetic Monte Carlo model used in this work, was described in detail previously.[4,16] We start by assuming a box with a distribution of sites of size 30×30×55 where the number of sites in the x- and y-directions is 30 and, in the z-direction is 55. The model is implemented on a cubic lattice with an inter-site distance of 1.8 nm, thus, the thickness of our simulated device is about 100 nm. In x- and y-directions periodic boundary conditions apply. The site energies are randomly distributed in a Gaussian distribution with $\sigma_{DOS}$ being the static disorder of the DOS and $E_0$ the mean energy

$$g(E) = \frac{1}{\sqrt{2\pi\sigma^2}} \exp\left[-\frac{(E - E_0)^2}{2\sigma_{DOS}^2}\right] \qquad \text{(S4)}$$

The hopping rates describe the hopping probability from the initial site, $i$, with energy, $E_i$, to the final site, $f$, with energy, $E_f$, in the above DOS is given by the Miller-Abrahams expression

$$\vartheta_{if} = \vartheta_0 \times \begin{cases} \exp\left(-\dfrac{E_f - E_i \mp q\overrightarrow{r_{if}} \cdot \vec{F} + \Delta E_C}{k_B T}\right), & E_j > E_i \\ 1, & E_j \leq E_i \end{cases} \qquad \text{(S5)}$$

Here, $\vartheta_0$ is the rate of a downward hop, which includes the exponential term describing the distance dependence, $F$ is the external electric field, and $\Delta E_C$ is the Coulomb energy variation calculated from interactions of all charges in the system and is given by $E_c = \frac{-q}{4\pi\epsilon_0\epsilon_r r_{eh}}$ ($\epsilon_r$ = 3.6 for organic materials) and $r_{eh}$ is the distance between two charge carriers. When two charges occupy the same site, the Coulomb term will diverge and in order to avoid the issue of double occupation, $E_c$ is truncated at the exciton binding energy, $E_B$, which is set to 0.5 eV.

Before a hop is made in the simulation, all of the hopping rates $\vartheta_{if}$ from current state $i$ to all possible states $f$ are calculated. Two uniformly distributed random numbers, $r_1$ and $r_2$, are generated where the first one lies between 0 and the sum of all hopping probabilities, $\sum \vartheta_{if}$ such that it chooses the transition from a set of possible hops $\{\vartheta_{if}\}$, which fulfils the conditions given below:

$$\sum_{j=1}^{\mu-1} \vartheta_{ij} < r_1 \leq \sum_{j=1}^{\mu} \vartheta_{ij} \qquad \text{(S6)}$$

where $\mu$ represents the hop that is made. Whereas the second random number should lie between 0 and 1, and this dictates the time passed between two hops and is represented in the program as

$$\tau = -\frac{\ln(r_2)}{\sum \vartheta_{ij}} \qquad \text{(S7)}$$

where $r_2$ is the randomly generated number between 0 and 1, and the denominator represents the sum of all transition rates. According to the Eq. S6, a specific hop is performed. For the next hops, the hopping probabilities are updated, and this process continues until a specified time has been reached, set such that a steady state is reached. The motion of charge carriers per unit area in any $(x, y, z)$ direction is calculated by summing over all the hopping movements in that direction, and the current density is determined by differentiating this with respect to time,

$$j_i(t) = \frac{d}{dt}\left(\frac{1}{A_i L_i}\sum_{\mu_i} q\Delta r_{\mu,i}\right) \qquad \text{(S8)}$$

where $A_i$ is the area of the cross-section of the box in the specific direction $i = x, y, z$ and $L_i$ is the corresponding length of the box. For direction perpendicular to the applied field, the current density obviously goes to zero, while the noise goes to the thermal limit, as discussed further in Supplementary Text S7 and Fig. S7b.

The morphology used for this work is called 'pillar', which is a representation of phase-separated donor-acceptor blend. Previously, this morphology was successfully used to reproduce the experimental jV-curves for TQ1:PCBM and PM6:Y6 solar cells with great accuracy.[16,18] This phase-separated morphology is achieved by having a $7 \times 7$ inclusion of acceptor material in a $10 \times 10$ unit cell of donor material in the x-y plane.

The input parameters for the simulations are given in Table S1 and were kept symmetric for holes and electrons.

**Table S1 Input parameters used for kMC simulations**

| Parameter [unit] | Value |
|---|---|
| Nearest neighbor distance [nm] | $1.8$ |
| Donor: LUMO & HOMO [eV] | $3.4; 5.2$ |
| Acceptor: LUMO & HOMO [eV] | $3.8; 5.6$ |
| Injection barrier [eV] | $0.2$ |
| Rate of downward hop, $\vartheta_0$ [$s^{-1}$] | $1 \times 10^{11}$ |
| Inverse exciton lifetime, $\vartheta_{ex}$ [$s^{-1}$] | $1 \times 10^{9}$ |
| Inverse CT lifetime, $\vartheta_{CT}$ [$s^{-1}$] | $3 \times 10^{7}$ |
| Exciton generation rate $G_0$ [$s^{-1}$ $m^{-3}$] | $3 \times 10^{28}$ |

The used parameters are typical values as found in our previous works and do not critically affect any of the results presented herein.[16,18,19]

## S7: Electronic Temperature Calculation in kMC

As in experiments, the kMC simulation also produces noise due to the hopping of charges in the system. The noise is only monitored when the system has achieved a steady state. The routine stores the motion of the charge in all three directions and calculates the time derivative of charges times displacements to get the current fluctuations using Eq. S8. The Fourier transform of this current signal is calculated while the DC value is subtracted. In a system with no contacts and an electric field in the z-direction, we were able to measure thermal noise in the direction perpendicular to the field, see Fig. S7b, while attaining shot noise in the direction of the field. Unfortunately, due to the presence of contacts and possible build-up of space charge next to the contacts, in a full device, the noise perpendicular to the field is dominated by other noise sources. In this work, we use the exact routine that we used experimentally to extract thermal noise from the kMC simulations.

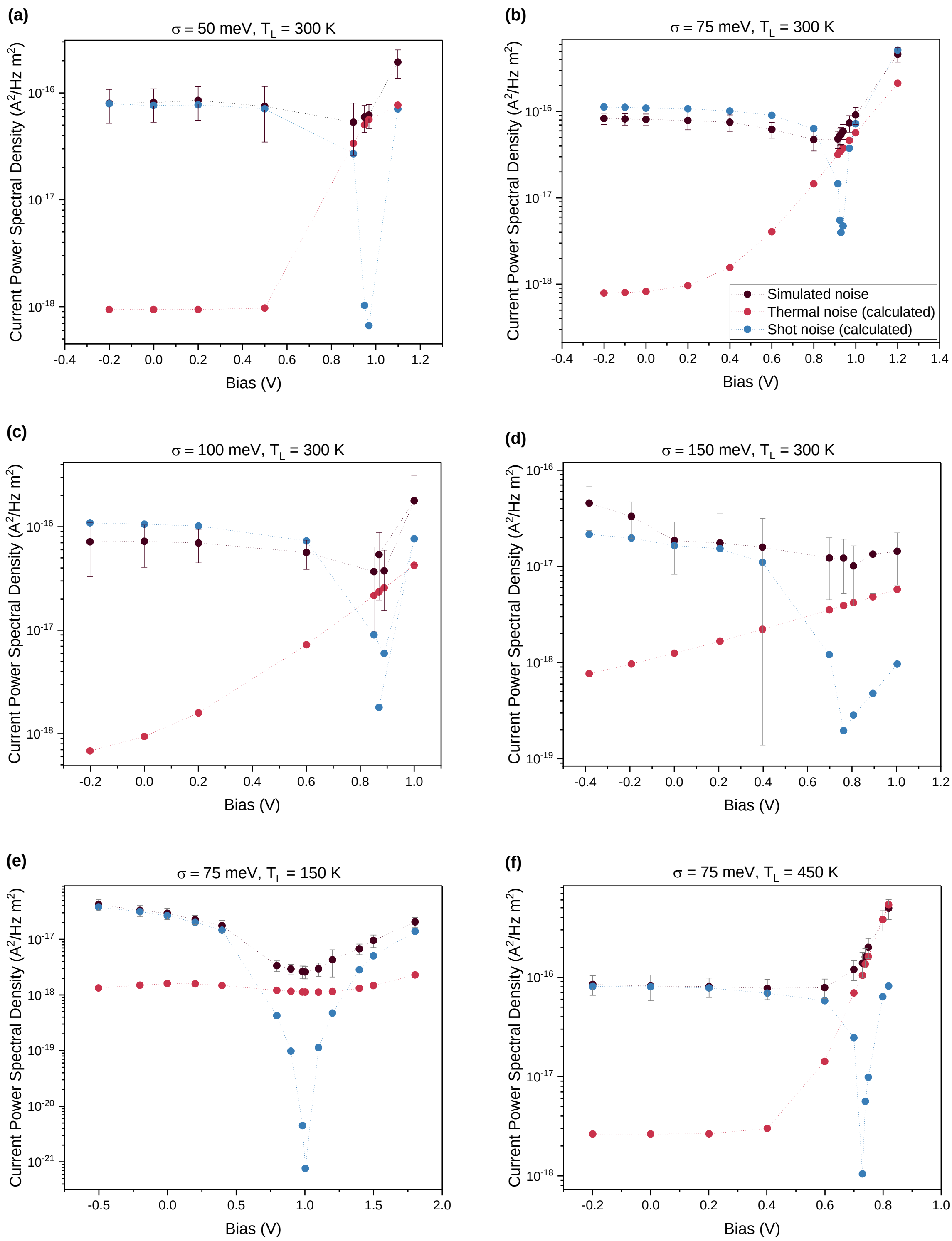

**Fig. S10 Simulated noise as a function of bias for different static disorder and lattice temperature.** Black, red and blue symbols are total simulated noise, calculated thermal noise and shot noise, respectively, with thermal and shot noise values calculated assuming the electronic temperature equals the lattice temperature and a Fano factor $F = 1$.

We start by simulating a solar cell with parameters outlined in Table S1. We apply different electric fields and simulate the output current and noise (Fig. S10). Analogous to the experimental findings, we observe noise predominantly characterized by shot noise at or near short-circuit conditions, and thermal noise at voltages close to open-circuit. It can be seen clearly in Fig. S10 that the shot noise with $F = 1$ surpasses the simulated noise. In simulations, since we have a better insight into the dynamics of charge carriers in the system, we attempt to estimate the Fano factor. We observe a correlation between the Fano factor and the fraction of charges escaping to the correct contact, called net escape and is defined as $(-J_{n,an} + J_{n,cat} + J_{p,an} - J_{p,cat})/J_{abs}$, where $J_{(n/p),(an/cat)}$ is the current density of photo-generated electrons/holes extracted via the anode/cathode and $J_{abs}$ is the current density corresponding to light absorption. We plot the absolute value of the net escape and $F$ calculated from $S_I/2qI$ in Fig. S11. The absolute value of the net escape goes close to zero around open-circuit conditions, which is in line with the understanding of the Fano factor (Supplementary Text S4).

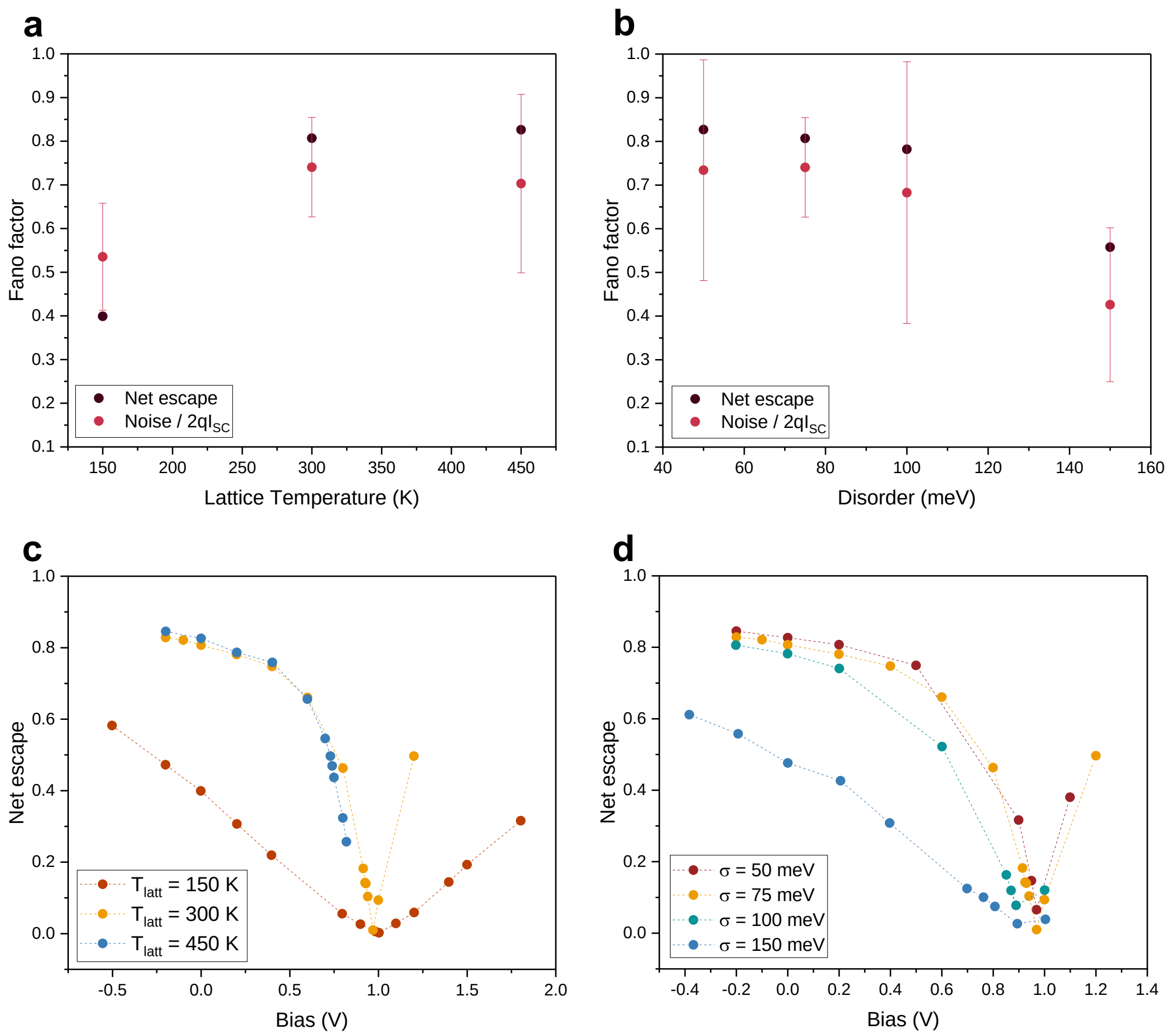

**Fig. S11 Comparison of Fano factor and extraction yield in kMC simulation.**
(**a**) and (**b**) The red dots are the Fano factor calculated using Eq. S2 for the same light intensity at which net escape was determined (black dots). (**c**) Variation of net escape with lattice temperature and (**d**) with disorder.

Thus, we use the net escape as Fano factor in our simulations to calculate the shot noise at open-circuit. The values range from 0.002 in the case of panel (e) to 0.5 for panel (f), and, as seen in Fig. S10, there is a strong effect of temperature on the Fano factor. We obtain the resistance at open-circuit and calculate the electronic temperature $T_{el}$ by subtracting the Fano factor-reduced shot noise.

$$T_{el} = \frac{Simulated\ Noise - F \times Shot\ Noise}{4k_B/R} \qquad (S9)$$

## S8: Noise Simulation of only Photo-generated Charges

Our kMC model allows to track the energy of photo-generated charges over time after dissociation from the parent exciton, which can be related to a temperature via

$$T_{photo}(t) = \frac{q}{k_B} * \frac{\sigma_{DOS}^2}{\left(E_0 - E_{photo}(t)\right)} \qquad \text{(S10)}$$

where $T_{photo}$ are the solid lines in Fig. S12a-c, $E_0$ and $\sigma_{DOS}$ are the center of DOS and static disorder of the same, $E_{photo}$ is the time-dependent mean energy of the population of charge carrier. Eq. S10 assumes that the (low density of) photo-generated charges are distributed thermally among themselves (cf. Ross/Nozik model) but not necessarily with a temperature equal to that of the lattice.

The simulations shown in Fig. S12 are performed for three different static disorders (50 meV, 75 meV and 100 meV) at open-circuit conditions. Previously, it was shown that the energy losses simulated by kMC are consistent with the experimental observations.[4] The scatter towards the end of the energy and temperature curves are due to poor statistics stemming from the fact that at longer timescales the number of charges left in the device is considerably lower. The extraction plots in Figs. S12d-f show the distribution of extracted charges on the left and the fraction of total extracted charges on the right. The extremely mobile, fastest 10%, of the total charges are extracted out of the system within 0.1 ns in a system with 50 meV disorder and 0.5 ns for 100 meV. Therefore, and since the maxima of the distribution of the extracted charges lies well after 50% of all charges are already extracted, determining a consistent mean temperature turned out to be impossible, and would probably not be meaningful due to the very dispersive nature of OPV.[19]

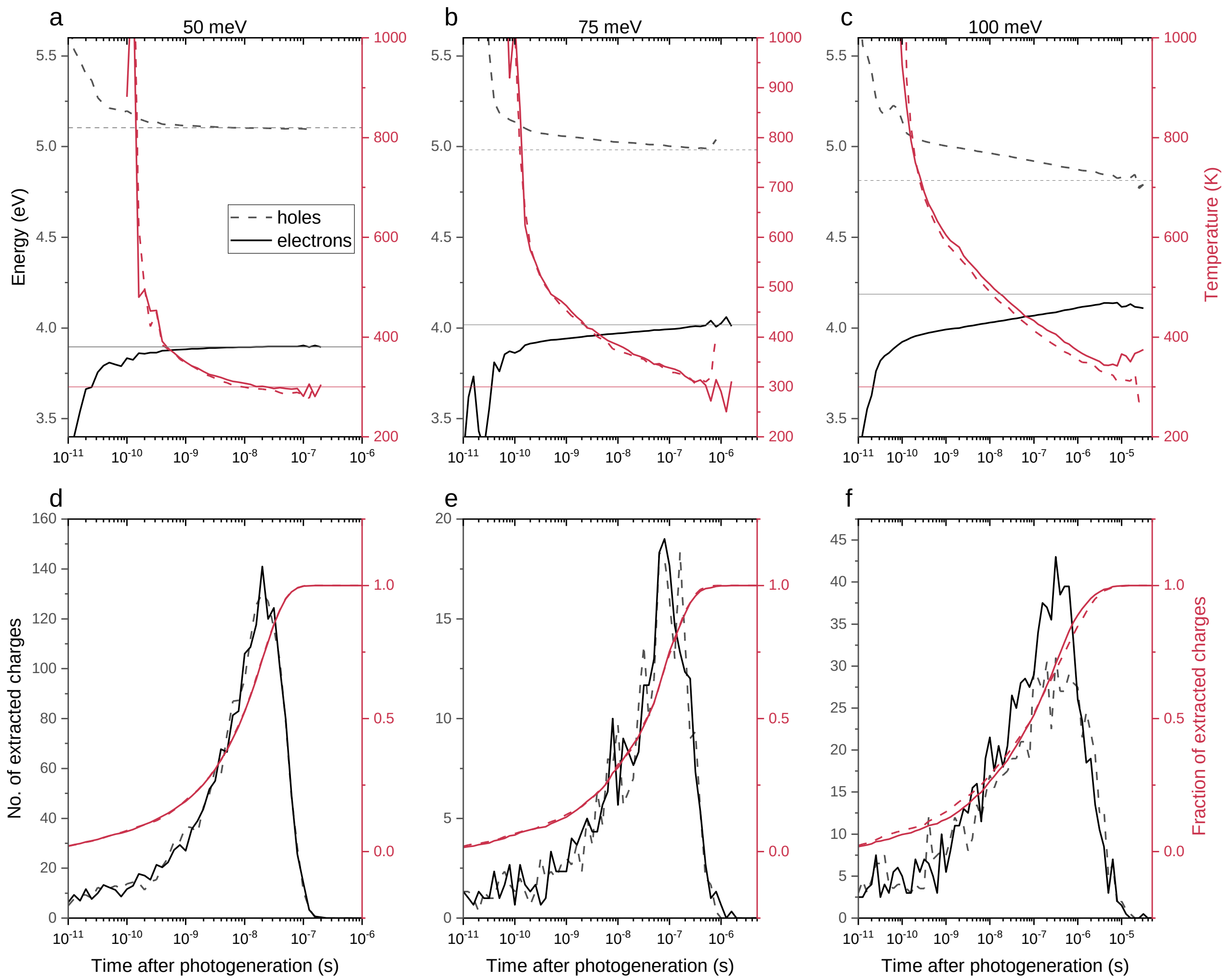

**Fig. S12 Thermalization loss and extraction fraction after photo-generation at open-circuit.** (**a**)-(**c**) Energy loss of photo-generated charges and the corresponding temperature for 50, 75 and 100 meV disorder at open-circuit condition. The grey horizontal lines correspond to the equilibrium energy, $\sigma_{DOS}^2/k_BT$ below (or, for holes, above) the center of the DOS and the red horizontal line indicates the lattice temperature of 300 K. (**d**)-(**f**) The extraction time distribution of photo-generated holes (dashed curve) and electrons (solid curve) and the corresponding integrated fraction of extracted charges as (solid and dashed) red curves as function of time after generation.

Figs. S12a-c differs slightly from the results shown previously,[16,18,20,21] due to a minor programming error (that only affected the calculated mean energy at long times and nothing else) being corrected. Although the energy of the photo-generated charges now reaches the equilibrium value (grey lines) at longer times after generation, this does not mean that the OPV at steady-state has equilibrated and our previous conclusions remain fully valid. The thermalization to equilibrium energy happens at much longer time scales than required for the extraction of the majority of the photo-generated charges. The histogram of extraction time peaks before full thermalization has been reached, which can be most clearly seen in the case of 75 and 100 meV disorder, but also in the 50 meV system the associated electronic temperature (red curve, calculated from Eq. S10) only reaches 300 K after more than 50% of the charges have been extracted. Note that even if a significantly larger fraction would have been extracted at the 'thermalization time', i.e., in the case of a hypothetically low disorder, the average temperature of all photo-generated charges in the system would still lie above the lattice temperature.

As argued in the main text, the photo-generated charges are extremely hot right after generation, but within the first few hops, they lose a significant part of their energy which can be seen in the steep decay of their temperature. In this particular simulation, the reduction is also due to the charges sitting next to the contacts getting extracted right after creation. The charges that attain complete thermalization are the charges that get stuck at the deep energy sites. From Fig. S13b, it can be seen that the photocurrent in OPV is carried by charges that possess high energy and not by charges that are sitting around the equilibrium energy.

## S9: Transport Energy from kMC

Knowing the exact value of the transport energy in our numerical simulations is essential as it is used to calculate the open-circuit voltage via the hot carrier solar cell model of Ross and Nozik.[22] An intuitive algorithm to determine the transport energy from numerical simulations was described by Oelerich et al.[23] The idea is to cut small energy intervals out of the DOS and observe for which energy range the charge carrier mobility is affected the most (see Fig. S13a). While this algorithm agrees nicely with analytical solutions for solvable systems, e.g. next-nearest neighbor hopping, it essentially requires running the same computationally expensive kMC simulations multiple times to get an exact result. Here, we will show that the transport energy can be extracted from our numerical results with less computational effort and that the determined values agree well with the algorithm by Oelerich et al.[23]

The kMC model tracks the charge displacement in the field direction for each hop, i.e. the distance a charge moves in field direction, as well as the mean energy of this hop, cf. Eq. S8. Dividing both values gives a time transient for the mean energy of charge displacement which in steady-state conditions and without contacts gives the same transport energy like the algorithm by Oelerich et al.[23] This methodology, however, can suffer from a similar issue like earlier algorithms that simply traced the most frequently visited energies.[24] If fast oscillations between at least three spatially close sites occur, they could dominate the statistics. This can be observed for simulations with contacts where a lot of charges quickly hop in and out of the device and if these charges have at least one intermediate hop to another energy level, the tracked energy of the oscillation does not average out completely. Without contacts this issue seems to be negligible for our specific simulations and the algorithm can be improved for simulations with contacts by tracking only the transport energy of photo-generated charges to circumvent the injection/extraction oscillation.

Another option to determine the transport energy is to calculate an energy-resolved current through the device and finding its maximum. Since the problematic oscillations of charges hopping in and out of the device happen at low energy states while the transport energy sits close to the center of the gaussian DOS, it is possible to remove the contribution of the contact oscillations. For this correction, it is necessary to run the simulations once with illumination and once in the dark to subtract the injection/extraction contributions (dotted red and black curve Fig. S13b respectively). The corrected data shows a peak with its maximum sitting exactly at the same energy as the transport energy of photo-generated determined via the mean energy of charge displacement as introduced above, see solid red curve in Fig. S13b. However, even after subtraction there is significant noise remaining from the charge injection and extraction hops. The results can be improved by looking at devices without injection (dotted orange curve) or by simulating hole- or electron-only simulations without contacts but comparable charge carrier densities (dashed blue curve) or by increasing the number of configurations to get better averages. While it is not directly obvious that simulations with and without injection as well as fully thermalized hole-only simulations result in the same transport energy, it can be seen in Fig. S13b that the transport energy is indeed independent of these changes for the systems studied in this work.

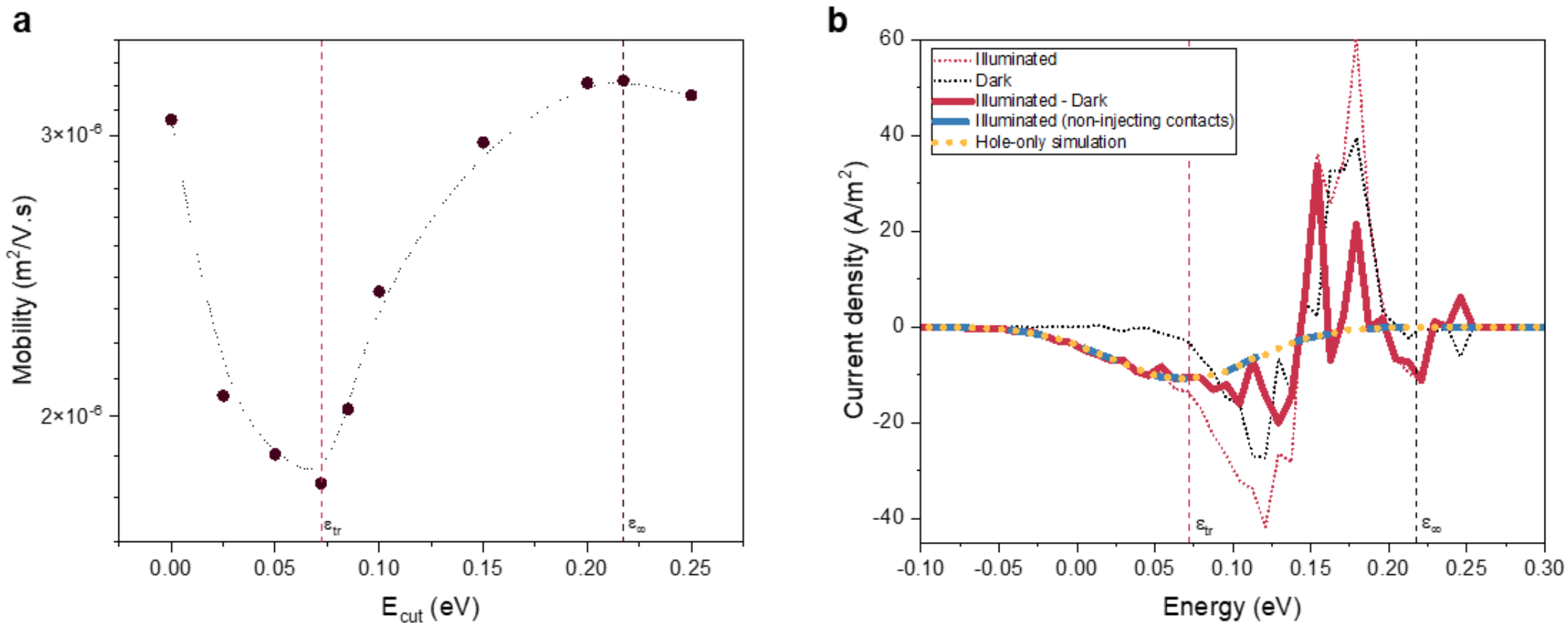

**Fig. S13 Determination of transport energy in simulation.**
(**a**) On cutting away energies at $E_{cut}$ with a width of 20 meV from the DOS, the change in mobility is determined. The minimum in the mobility matches with the transport energy (dashed red vertical line) determined by our kMC model. The simulation is carried out for 75 meV and 300 K lattice temperature and the center of Gaussian DOS sits at 0 eV. (**b**) The energy-resolved current density for an OSC with the same parameters as given in Table S1 at short-circuit conditions. Note that the peak in the curve obtained from the difference the illuminated and dark curves (red curve) matches with the OSC with non-injecting contacts and hole-only simulations. The vertical dashed red line is the transport energy used in Eq. 1 and matches with the transport energy determined using different routines. The vertical dashed black line is the equilibrium energy.

With respect to the energy filter required for working hot carrier solar cells, it is important to mention that it was confirmed that the extraction energies at the contacts and the energy-resolved current through the device and thus transport energy follow exactly the same distribution.

## S10: Reciprocity Analysis for Determining the Radiative Limit of $V_{OC}$

The quasi-Fermi level splitting (QFLS) without the loss channels is equivalent to the radiative limit of the $V_{OC}$.[25]

$$eV_{OC}^{rad} = k_B T \, ln\left(\frac{J_{SC}}{J_0} + 1\right) \qquad \text{(S11)}$$

The reverse dark saturation current is obtained by calculating the overlap of the external quantum efficiency (EQE) of the OPV with the blackbody spectrum,[26]

$$J_0 = q \int_0^{\infty} EQE_{PV}(E) \, \varphi_{BB}(E) \, dE \qquad \text{(S12)}$$

Where the blackbody spectrum is given by

$$\varphi_{BB}(E) = \frac{2\pi E^2}{h^3 c^2} \frac{1}{e^{\frac{E}{k_B T}} - 1} \qquad \text{(S13)}$$

Here, the EQE for the OPV is

$$EQE_{PV}(E) = IQE_{PV}(E)\varphi_{abs}(E) \qquad \text{(S14)}$$

where $IQE_{PV}$ is the internal quantum efficiency which we can set to unity for the calculations and $\varphi_{abs}$ is the absorption spectrum. For the simulations undertaken in this work, both acceptor and donor are set to absorb equally. We only consider the absorption in the CT and $S_1$ states of both materials as the higher absorption is cut-off by the steep blackbody spectra.

$$\varphi_{abs}(E) = a \, \varphi_{CT}(E) + \varphi_{S_1}^{donor} + \varphi_{S_1}^{acceptor} \qquad \text{(S15)}$$

$$a = \frac{\vartheta_{CT}}{\vartheta_{S_1}} * n \qquad \text{(S16)}$$

where $\vartheta_{CT} = 3 \times 10^7 \, s^{-1}$ and $\vartheta_{S_1} = 1 \times 10^9 \, s^{-1}$ are the CT and exciton recombination rates, cf. Table S1. The factor $n$ in Eq. S16 accounts for unequal number of sites available for CT and $S_1$ absorption in the simulation. For the 'pillar' morphology used in this work, we have a $7 \times 7$ inclusions of acceptor material in $10 \times 10$ unit cell of donor material. The CT states are at the interface of the two, thus, the ratio between CT and $S_1$ state, $n = 0.28$. Using this analysis, we obtain the radiative limit of open-circuit voltage for kMC simulations (see Fig. S14), which is close to the QFLS.

In experiments, it is not as straightforward to calculate the dark saturation current $J_0$ as it is in simulation. Although with sensitive EQE measurements, one is able to probe deeper into the DOS, it is still limited by the noise from the measurement system, biasing source, light source as well as the DUT's thermal and shot noise.[27] Consequently, discrepancies arise on determining the lower limit of integral in Eq. S12. In Fig. S14, we show that upon changing the lower limit one can obtain rather different $V_{OC}^{rad}$. In panel (b), a gaussian is fit at the low energy tail states of EQE while panel (c) shows an exponential extrapolation. The calculations are performed for three different temperatures, one at room temperature 300 K, while others at electronic temperatures obtained from noise spectroscopy for PM6:Y6 for different Fano factors. While we are aware that the employed reciprocity relation is inadequate in conjunction with the far-from-equilibrium picture that we describe in this work, we anyway calculate $V_{OC}^{rad}$ as a lowest order approximation.

Irrespective of the low-energy extrapolation, the curves in Fig. S14b,c show that the radiative limit of the open-circuit voltage depends critically on both the assumed temperature and the lower integration limit. In the near-equilibrium picture, the blue curve (300 K) is typically cut off at the upper plateau, such that the measured value lies below it, which is then attributed to various additional losses.[28,29] Using a more realistic electronic temperature makes $V_{OC}^{rad}$ more or less integration limit independent, and sitting below the measured $V_{OC}$ value.

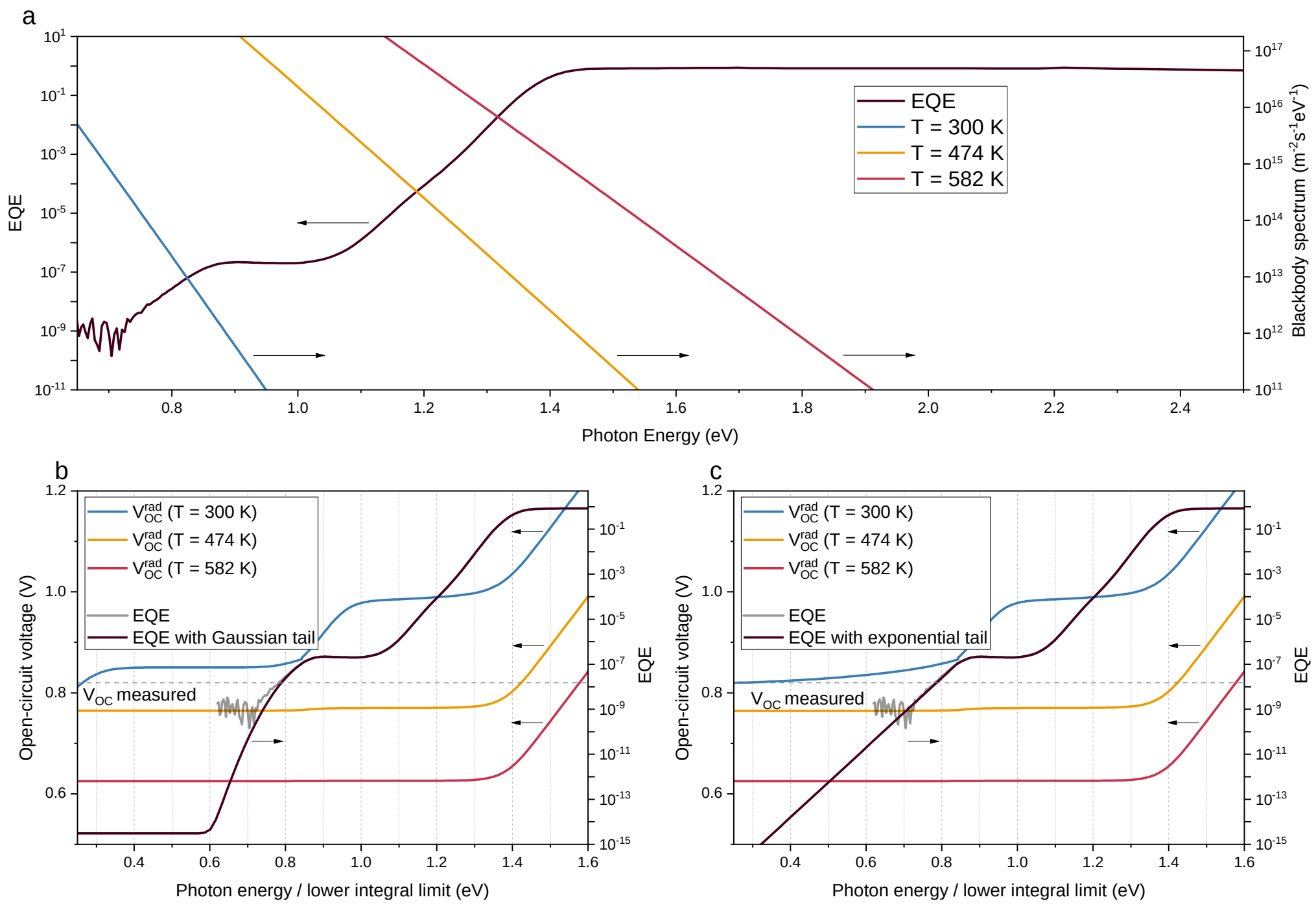

**Fig. S14 Determination $V_{OC}^{rad}$ from EQE measurements.**
(**a**) EQE measurement of PM6:Y6 OSC is shown by the grey curve. The black curve is obtained by fitting the low energy tails with different distributions. (**b**) Tail state of EQE fitted with a Gaussian. The colored lines are $V_{OC}^{rad}$ determined from Eq. S11 by using $T = 300$ K, 474 K and 582 K. (**c**) Tail state of EQE fitted with an exponential and corresponding $V_{OC}^{rad}$.

## S11: Determining Non-Equilibrium $V_{OC}$ using the Radiative Limit of $V_{OC}$ from the Reciprocity Relation

The open-circuit voltage in the Ross/Nozik framework (Eq. 1 in the main text) has a dependence on $\Delta\mu$ which is the difference in quasi-Fermi level of electrons and holes, which determines the $V_{OC}$ of the solar cell in absence of hot-carrier effects. Experimentally, it is extremely tedious to predict this QFLS, but instead, the upper (radiative) limit of $V_{OC}$ can be more easily determined by the reciprocity relation (Eq. S11). Hence, we calculated this radiative limit as described in Supplementary Text S10 for different lattice temperatures and disorders for the parameters used in the kMC simulations.

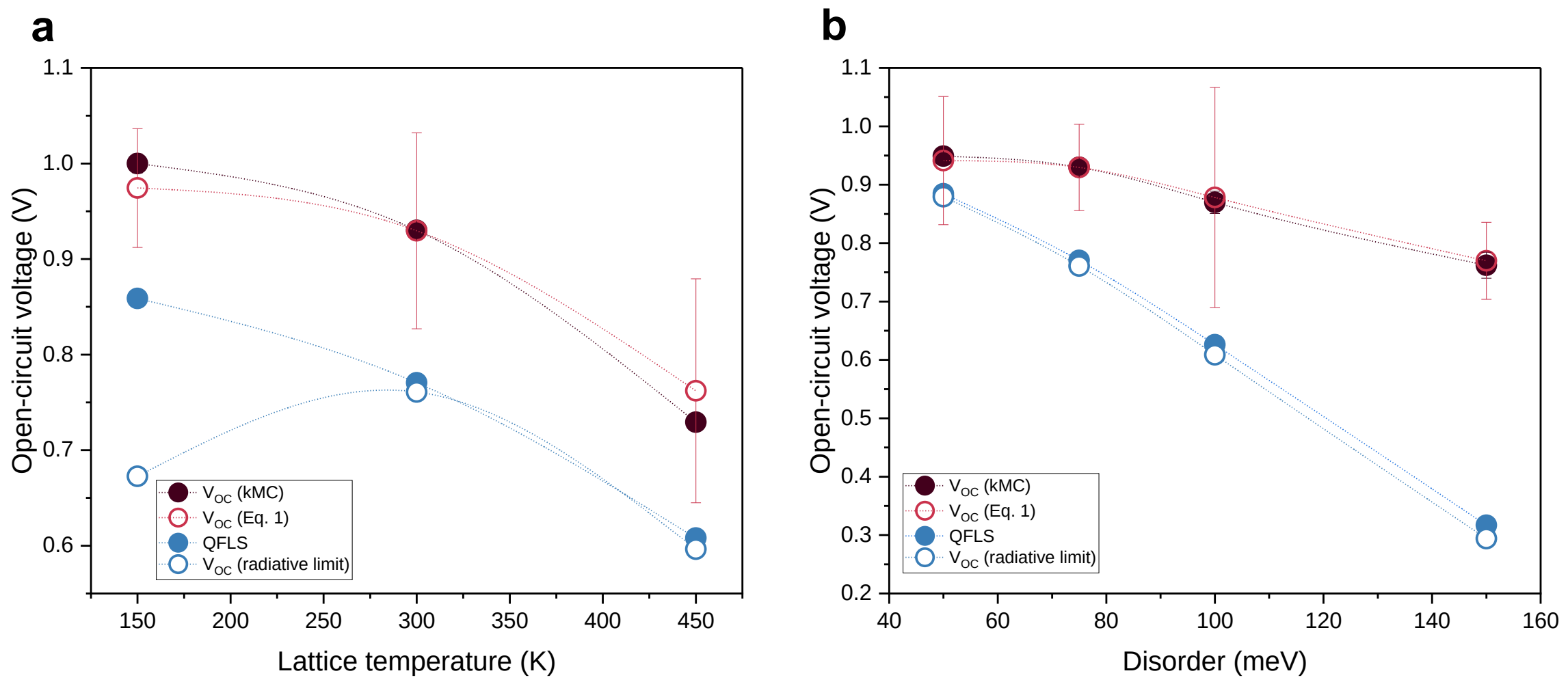

**Fig. S15 Non-equilibrium $V_{OC}$ from Eq. 1 using the radiative limit.**
(**a**) For different lattice temperature and a constant static disorder of 75 meV and (**b**) for different disorder and constant lattice temperature of 300 K, the $V_{OC}$ calculated using Eq. 1 (open red symbols) agrees well with the $V_{OC}$ obtained from kMC (filled black symbols) with using the radiative limit (open blue) determined from Eq. S11 instead of the quasi-Fermi level splitting (closed blue symbols) as starting points.

## S12: Non-equilibrium $V_{OC}$ Calculation from Seebeck Coefficient

In the main text, the equation for the non-equilibrium $V_{OC}$ was rewritten to emphasize that voltage gains resulting from the hot-carrier effect arise from the Seebeck effect between the hot photo-generated charge carriers and the cold electrodes (Eq. 3). The Seebeck coefficient in the kMC simulations is calculated via

$$S = \frac{\varepsilon_F - \varepsilon_{tr}}{T_{el}} \qquad (S17)$$

and has a dependence on DOS occupancy in the range of $10^{-5}$ to $10^{-3}$ ($10^{15}$ – $10^{17}$ $cm^{-3}$) (see Fig. S4). Since, in our simulations, electrons and holes are equivalent, we determine the Seebeck coefficient, $S$, from electron-only simulations. Plugging in the resulting values of $S$ in Eq. 3, we obtain a very good agreement of non-equilibrium $V_{OC}$ (Eq. 3) and $V_{OC}$ from kMC as shown in Fig. S16.

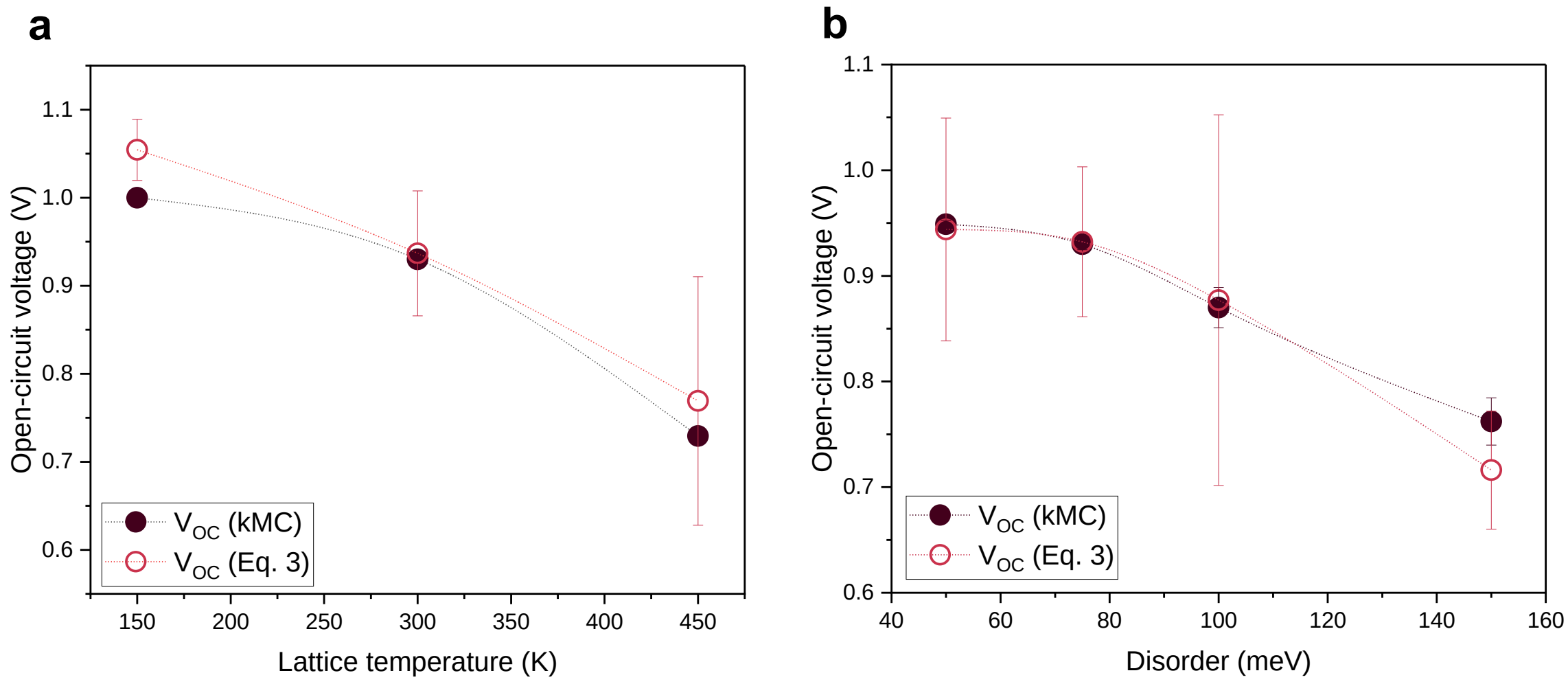

**Fig. S16 Non-equilibrium $V_{OC}$ calculated from Eq. 3.**
(**a**) For different lattice temperature and a constant static disorder of 75 meV and (**b**) for different disorder and constant lattice temperature of 300 K, the $V_{OC}$ calculated using Eq. 3 (open red symbols) agrees well with the $V_{OC}$ obtained from kMC (filled black symbols).

## S13: Spatial Evolution of Energy of Photo-generated Charges as a Function of Initial (Excess) Energy

The simulations used to observe the energy of photo-generated charges as a function of travel distance (as depicted in Fig.4d) is set up differently compared to the other simulations presented in this work. The simulations were carried out for a bilayer and a single bound electron-hole pair (CT state) is generated at the interface of this bilayer, with a specified excitation energy. The dielectric constant is artificially increased to a very high value in order to suppress the Coulomb interaction between the two charges. Each simulation generates only one CT state at a time, effectively eliminating other processes such as recombination. Thus, the fate of the charge is to be either trapped in the device or to be extracted. The simulations are averaged for 500 individual runs. The device thickness is varied and with it does the extraction time. The energies of the photo-generated charges after 10% and 50% of charges are extracted and plotted as a function of distance between the generation of CT state and the extraction point (Fig. 4d).

The charges with extremely high excess energy, i.e. the ones created in the top-half of the DOS, lose most of their energy in the first hop, after which it is equivalent to the charge that is generated at the center of the DOS. Without recombination, charges generated close to or below equilibrium energy need to climb up to the transport energy in order to be extracted. The purpose of these simulations was to single out how excess energy influences, and is lost during the charge transport.

It should be noted that the results as depicted in Fig. 4d cannot be directly mapped onto a solar cell under steady-state illumination, as state filling effects are neglected and many low-lying charges would recombine before they rise to transport energy. It was shown by Upreti et al. that the $V_{OC}$ remains constant with increasing thickness and that hot-carrier effects are still pronounced.[16]

## S14: Hot Narrow Band Absorber OSC

As mentioned in the main text and in Supplementary Text S13 above, excitations to energies above the center of the DOS are lost to thermalization within the first few hops. The same issue appears after charge transfer at the interface between donor and acceptor, when charges fall into the upper half of the DOS. Therefore, the optimal hot carrier solar cell has a similar energy for the first excited singlet state ($S_1$) and the charge transfer (CT) state. We implement this kind of system in our kMC simulation by taking the parameters defined in Supplementary Text S6, but with reduced bandgaps such that a small driving force $E_{S1} - E_{CT} = 0.05$ eV for charge transfer remains, which requires $HOMO_{donor} - HOMO_{acceptor} = LUMO_{donor} - LUMO_{acceptor} \approx 0.3$ eV.[30] In the following, we compare the power conversion efficiency for a system with full illumination, narrow band illumination around the center of the Gaussian DOS and narrow band illumination around the transport energy.

The necessary input power density $P_{\text{in}}$ for a narrow band absorption can be easily derived for a delta-distributed photon flux around the central energy $E_{\text{excite}}$ and is given by

$$P_{\text{in}} = qGLE_{\text{excite}} = j_{SC}\, E_{\text{excite}} \qquad \text{(S18)}$$

where $q$ is the elementary charge, $G$ is the generation rate and $L$ is the device thickness. If we assume $EQE = 1$, this calculation can be extended to a Gaussian-shaped absorber and yields the same equation for the power input for symmetric excitation of finite width $\Delta E_{\text{excite}}$ around the center of the Gaussian. It is possible to shift $E_{\text{excite}}$ with respect to the center of the Gaussian by applying additional correction terms that consider the modified absorption profile, but it should be kept in mind that the assumption $IQE = 1$ becomes unreasonable when $E_{\text{excite}}$ is pushed too far towards the low-energy tail of the Gaussian. Fig. S17 shows the results for full illumination, narrow band excitation around the center of the Gaussian and for excitation around the transport energy. The short-circuit current for narrow band absorption is naturally lower as limiting the absorption window to a specific photon energy reduces the generation rate $G$. Rescaling the simulated currents by $G_0/G$, where $G_0$ is the generation rate for full illumination, leads to matching curves for full illumination and narrow band illumination around the center of the Gaussian. Consequently, the generated power is the same if the same number of photons are absorbed but the power conversion efficiency scales by

$$\text{PCE}_{\text{narrow}} = \text{PCE}_{\text{Full}} \frac{P_{\text{In,Full}}}{j_{SC}\, E_{\text{excite}}} \qquad \text{(S19)}$$

Applying this scaling to PM6:Y6 with $\text{PCE}_{\text{Full}}$ = 15%, $j_{SC} = 260$ A/m$^2$ at AM1.5 (=1000 W/m$^2$) and a $S_1$ absorption maximum at ~1.4 eV yields $\text{PCE}_{\text{narrow}} = 41.2\%$.

For the excitation around the transport energy with $G_0$, the open-circuit voltage is slightly reduced but the generated power does not change due to a small improvement in fill factor. The reduced photon energy, however, reduces the input power by another 6% which applied to the PM6:Y6 scenario improves the power conversion efficiency to 43.9%. Most likely this value should be regarded as an upper limit, since the assumption $EQE = 1$ might not be strictly correct below the absorption maximum, but the simulation remains insightful as it seems reasonable to expect a maximum PCE between the absorption maximum and the transport energy due to the competing effects of absorption efficiency and thermalization losses down to transport energy.

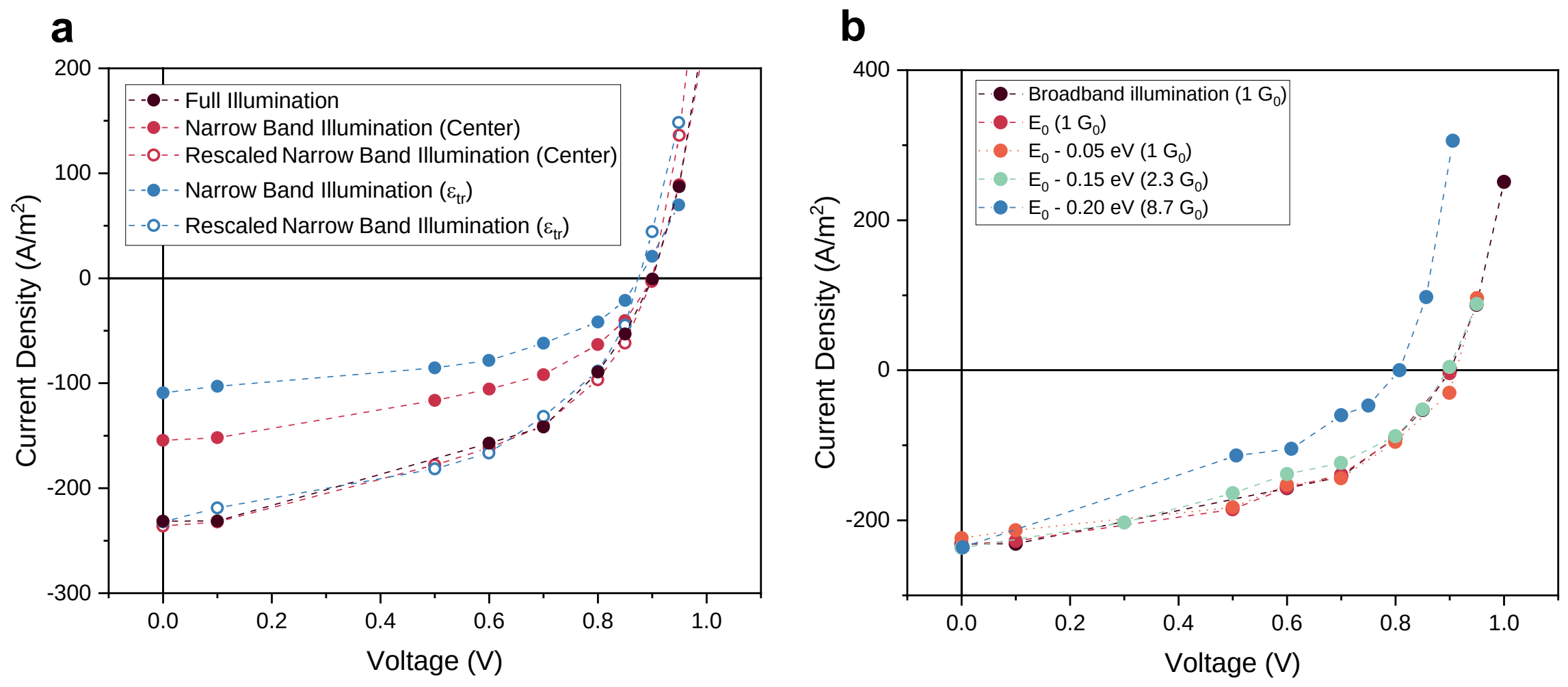

**Fig. S17 Narrow band absorber hot carrier solar cell.**
(**a**) Current-voltage curves for kMC simulations considering full illumination, narrow band illumination around the center of the absorption Gaussian and narrow band illumination around the transport energy. The energy window for narrow band illumination has a width of 0.1 eV and for rescaling the currents are multiplied by $G_0/G$, where $G_0$ is the generation rate for full illumination. (**b**) Current-voltage curves for kMC simulations considering full (black symbols) and narrow band illumination at and below $E_0$, the center of the DOS.

As a follow-up to Fig. 4d in the main text, we tested the effect of the excitation energy on the open-circuit voltage. While being an extremely challenging experiment to perform due to the weak absorption below the bandgap, it is numerically possible provided a sufficiently large number of absorbing sites is present in the kMC simulation box. Using the same energy landscape described above, that is, with a small driving force $E_{S1} - E_{CT} = 0.05$ eV, we simulate IV curves under varying photon energies, such that photo-generated charges are created only at specific energy sites like at the center of DOS, $E_0$, or at energies lower than $E_0$ including at equilibrium energy which is at $E_0 -$ 0.2 eV, where they possess no excess energy.

To rule out that $V_{OC}$ is affected by reduced short-circuit currents, we adjusted the illumination accordingly. This is necessary to compensate for increasing recombination losses at low excitation energies because excitons generated deep in the tail of DOS, cf. blue and green symbols in Fig. S17b, are mostly immobile due to the limited number of energetically accessible sites around them and cannot efficiently dissociate into free charges. For the extreme case of excitation 0.20 eV below the DOS center, blue symbols, the generation rate must be increased to $8.7 \times G_0$ to get similar short-circuit currents as for broadband illumination. These results confirm those by van Eersel et al. that some excess energy, defined with respect to the equilibrium energy, is needed for efficient free charge generation by allowing charges to thermalize in the DOS while climbing up the Coulomb potential.[30] Hence, charges generated sufficiently far above the equilibrium energy to allow for more or less efficient free charge generation, orange and green symbols, begin with an excess energy that is also sufficient to see hot-carrier effects which in turn give rise to a similar $V_{OC}$ as the broadband illumination (within the limits of our numerical resolution). Meanwhile, charges generated (too) close to the equilibrium energy, blue symbols, suffer from a reduced IQE and a reduced $V_{OC}$.